\documentclass[aps,prb,twocolumn,showpacs,superscriptaddress]{revtex4-1} 

\usepackage[dvips]{graphicx}
\usepackage{color}
\usepackage{psfrag}
\usepackage{ulem}
\usepackage{bbm}

\usepackage{amsthm}
\usepackage{amsmath}
\usepackage{amsfonts}
\usepackage{amssymb}

\usepackage[utf8]{inputenc}
\usepackage{mathptmx}
\usepackage{t1enc}
\usepackage[margin=0.7in]{geometry}

\usepackage{graphicx}
\usepackage{bbold}

\newcommand{\be}{\begin{equation}}
\newcommand{\ee}{\end{equation}}

\begin{document}

\title{Cluster Truncated Wigner Approximation in Strongly Interacting Systems}
\author{Jonathan Wurtz}
\affiliation{Department of Physics, Boston University, 590 Commonwealth Ave., Boston, MA 02215, USA}
\author{Anatoli Polkovnikov}
\affiliation{Department of Physics, Boston University, 590 Commonwealth Ave., Boston, MA 02215, USA}
\author{Dries Sels}
\affiliation{Department of Physics, Boston University, 590 Commonwealth Ave., Boston, MA 02215, USA}
\affiliation{Department of Physics, Harvard University, 17 Oxford st., Cambridge, MA 02138, USA}
\affiliation{Theory of quantum and complex systems, Universiteit Antwerpen, B-2610 Antwerpen, Belgium}
	
\begin{abstract}
We present a general method by which linear quantum Hamiltonian dynamics with exponentially many degrees of freedom is replaced by approximate classical nonlinear dynamics with the number of degrees of freedom (phase space dimensionality) scaling polynomially in the system size. This method is based on generalization of the truncated Wigner approximation (TWA) to a higher dimensional phase space, where phase space variables are associated with a complete set of quantum operators spanning finite size clusters. The method becomes asymptotically exact with the increasing cluster size. The crucial feature of TWA is fluctuating initial conditions, which we approximate by a Gaussian distribution. We show that such fluctuations dramatically increase accuracy of TWA over traditional cluster mean field approximations. In this way we can treat on equal footing quantum and thermal fluctuations as well as compute entanglement and various equal and non-equal time correlation functions. The main limitation of the method is exponential scaling of the phase space dimensionality with the cluster size, which can be significantly reduced by using the language of Schwinger bosons and can likely be further reduced by truncating the local Hilbert space variables. We demonstrate the power of this method analyzing dynamics in various spin chains with and without disorder and show that we can capture such phenomena as long time hydrodynamic relaxation, many-body localization and the ballistic spread of entanglement.
\end{abstract}

	\maketitle

\section{Introduction}
Phase space methods recently emerged as very powerful tools for simulating dynamics of quantum systems typically close to some classical limit. Such methods found a wide range of applications in many areas of science such as quantum chemistry, quantum optics, atomic systems, quantum chaos, condensed matter physics and others. Generally these methods replace exponential complexity of simulating quantum systems with the power law complexity of classical systems. This comes at the price of the classical dynamics being nonlinear, leading to potential instabilities. Quantum effects introduce additional stochastic dynamical noise on classical equations of motion, which is usually described by non-positive probability distribution and leads to rapidly increasing complexity of exact simulations with time. A common approximation, where this quantum noise is neglected, is known as the truncated Wigner approximation (TWA)\cite{Hillery1984, Steel1998, Blakie2008, Polkovnikov2010}. At the same time TWA accounts for the initial noise encoded in the Wigner function describing the initial state. It is this noise which makes TWA distinct from mean field approaches. A standard limitation of TWA is that it is usually applicable for a finite time, which is set by the effective Planck's constant and by the magnitude of nonlinearities (interactions) in the system. TWA becomes exact either in the classical, or the noninteracting limit \cite{Polkovnikov2010}.

In strongly interacting systems far from the classical limit standard phase space methods do not usually lead to simplifications or may even result in uncontrolled approximations. However, as was demonstrated in Refs.~\onlinecite{Davidson2015, Davidson2017} one can partially circumvent these difficulties by increasing dimensionality of phase space by adding extra/(hidden) degrees of freedom representing correlations and treating them as independent variables. Thus in Ref.~\onlinecite{Davidson2015} a set of three level (spin one) systems was efficiently treated by first mapping them to SU(3) representation with 8 classical degrees of freedom per spin and then applying SU(3) TWA instead of the traditional TWA based on the SU(2) spin representation \cite{Polkovnikov2010, Orioli2017} with 2 degrees of freedom per spin. This increased phase space dimensionality allows one to map all local on-site interactions to an effective on-site magnetic field, linearizing the local part of the Hamiltonian. Thus the only source of non-linearity potentially leading to errors due to TWA come from the inter-spin interactions. As was demonstrated in Ref.~\onlinecite{Davidson2015} using the $SU(3)$ representation significantly improves the accuracy of TWA. Similarly, TWA was adopted to fermion bilinears, which can be identified as classical phase space variables~\cite{Davidson2017}. These bilinears (or string variables) form a closed SO($2N$) Lie-algebra for $N$-single particle orbitals, which can be described by a phase space of dimensionality $2N^2$ instead of the naive dimensionality of $2N$. In Refs.~\onlinecite{Davidson2017,Schmitt2017} it was shown that fermion TWA accurately describes quantum dynamics in various non-trivial setups like a two-channel model, expansion of interacting fermions and dynamics of the SYK model.

In this work we extend the ideas of Ref.~\onlinecite{Davidson2015} by developing a cluster TWA (CTWA) approach, where we treat all operator degrees of freedom confined to a cluster as independent phase space variables. In this way we first map quantum Hamiltonians to large $\mathcal N$-models, where $\mathcal N=D^2$ and $D$ is the Hilbert space size of the cluster. Then dynamics of the system is governed by the classical equations of motion as in standard large-$\mathcal N$ theories (Ref.~\onlinecite{Yaffe1982}), with an additional ingredient that the initial conditions are sampled according to a Gaussian distribution set by the initial density matrix of the system. By construction, CTWA becomes asymptotically exact as we increase the cluster size and thus gives a controllable expansion for simulating dynamics in quantum systems, at least in principle. As it will be clear from our discussion, mathematically clusters can be defined in an arbitrary way by splitting the Hilbert space into a sum of orthogonal subspaces (clusters) and mapping the operator basis spanning these subspaces (which form a closed SU(D) algebra) into classical degrees of freedom.   In this work we will focus into local spatial clusters.

\begin{figure}
  	\includegraphics[width=\linewidth]{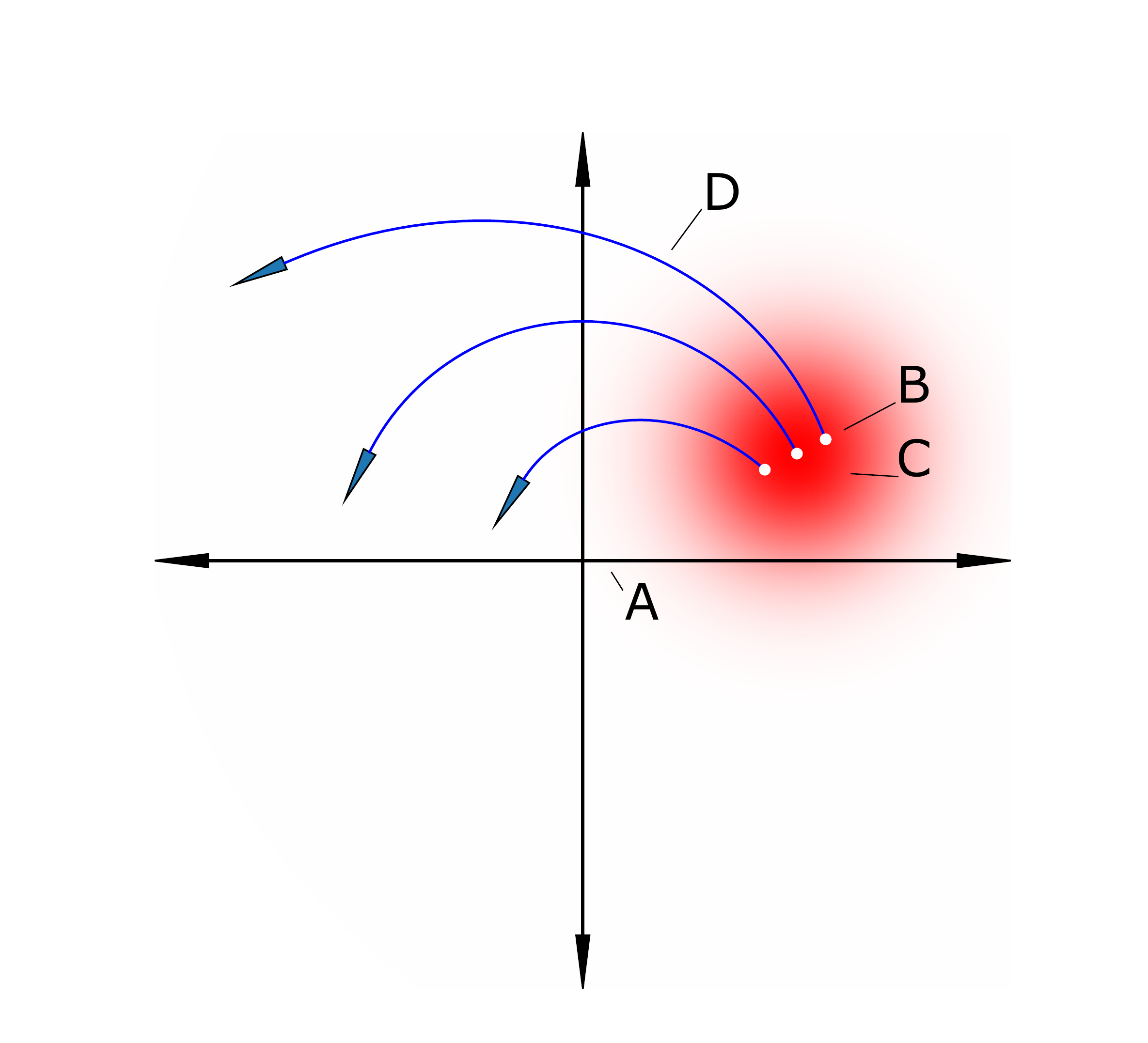}
  	\caption{Visual example of CTWA evolution in phase space. The coordinate system (A) (Black lines) parameterizes a phase space for points $\{x\}$ and functions $O(\{x\})$ mapped from operators $\hat O$. Points in phase space (B) (white dots) are selected from a probability distribution (C) (red blob) and evolved independently according to mean field classical equations of motion (D) (blue traces). As evolution is nonlinear, nearby points diverge in time. Observables are found by averaging over all sampled points.}\label{fig:phasespace_example}
  \end{figure}

One of the advantages of using the phase space formalism is that it allows one to discuss equations of motion, observables, time correlators, and initial conditions which include both quantum and thermal fluctuations. On the formal level, TWA can be derived by projecting exact time evolution of the density matrix into an appropriate operator product subspace. In this sense TWA is very similar to mean field or the time-dependent variational principle (TDVP)~\cite{Kramer2008,Leviatan2017, Shi2017}. However, because the initial Wigner function can never be represented as a single point in this space (due to e.g. the uncertainty principle) TWA necessarily involves a statistical mixture of different classical trajectories (see Fig.~\ref{fig:phasespace_example}) over a range of initial conditions in phase space, which generally cannot be expressed as a single mean field evolution.

 This statistical mixture in phase space allows one to simulate build up of quantum correlations and entanglement from initial product states, which is encoded in the classical mutual information induced by nonlinear evolution in phase space. Let us point that the main limitation of the cluster TWA is rapid scaling of the phase space dimensionality with the cluster size. E.g. for spin one-half systems the phase-space dimensionality of a cluster of size $N$ scales as $D^2=4^N$. We show that this number can be reduced to $D=2^N$ by employing multiple conservation laws analogous to spin conservation in the standard $SU(2)$ case. Equivalently one can understand this dimensional reduction using the classical analogues of the Schwinger boson representation of the phase space variables.  On a formal level these dimensional reductions map classical (mean field) equations for density functions to classical (mean field) equations for the wave functions. At the end of the paper, following the ideas developed in Ref.~\onlinecite{Leviatan2017} for mean field dynamics, we comment how phase space dimensionality can be reduced further by using incomplete local Hilbert space constructed from e.g. the matrix product state operator basis~ \cite{Verstraete2009, Orus2014, Pekker2017}.

The rest of this paper will be structured as follows. Sections \ref{sec:operator_twa} and \ref{sec:SB_TWA} cover the outline of the CTWA for the two methods: Operator, and Schwinger Boson CTWA. Section \ref{sec:CTWA_example} gives a simple example. Sections \ref{sec:Ising_results} and \ref{sec:1d_results}, will give various nontrivial demonstrations of the method. Section \ref{sec:GaussCTWA} outlines details of the method, including rationale for the Wigner function and dimensional reduction of the operator CTWA.

\section{Cluster TWA: operational manual}\label{sec:operational_manual}

Postponing a detailed discussion of the cluster TWA until the end of the paper, let us start with a brief summary of the method and demonstrating how it can be applied to a simple example. For illustration of the method we will use a system of  interacting spins ${1\over2}$ described by the Hamiltonian:

\begin{equation}\label{eq:hamiltonian}
  \hat H = \sum_{ij} J_{ab}^{ij}\hat \sigma_a^{(i)}\hat \sigma^{(j)}_b + \sum_j B_a^{j} \hat \sigma_a^{(j)},
\end{equation}
where $a,b\in x,y,z$ denote the indices of Pauli matrices and $i,j$ denote different spins on the lattice. The couplings $J_{ab}^{ij}$ can be either short range of long range, and both  $J_{ab}^{ij}$ and $B_a^j$ can generally be time-dependent.

\begin{figure}
	\includegraphics[width=\linewidth]{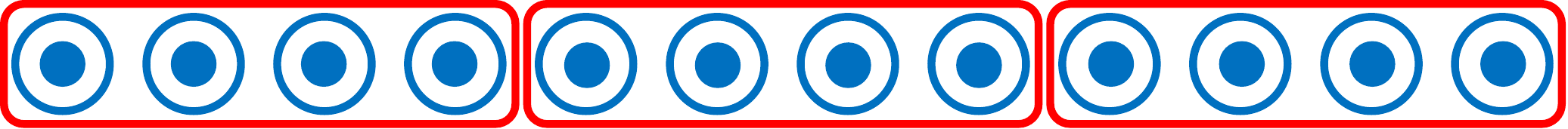}
	\caption{\textbf{Example of clustering of spins}. Blue circles are physical spins $1\over 2$, while red rectangles are the clusters. Hamiltonian dynamics within each cluster is exact, while interactions between clusters are treated approximately.}\label{fig:cluster_example}
\end{figure}

\subsection{Operator Cluster TWA}
\label{sec:operator_twa}

Here we describe the necessary steps for implementing the operator CTWA.

\begin{enumerate}

\item Split the system into clusters labeled by indices $i',j'$, e.g. like shown in Fig.~\ref{fig:cluster_example}. Define a complete operator basis $\{\hat X^{i'}_\alpha\}$, $\alpha=0,\dots D^2-1$ spanning the Hilbert space of each cluster ${i'}$, where $D$ is the Hilbert space dimension of that cluster: $D=2^N$ for a cluster of $N$ spins ${1\over 2}$. 
Two examples of a complete basis are products of Pauli matrices (section \ref{sec:operatorclusterTWA}) and all rank-1 Hermitian matrices (section \ref{sec:general_ICs}). We will use the convention that $\hat X_0^{i'}=\hat {\bf I}$ is the identity operator and all the remaining operators $\hat X_\alpha^{i'}$ are traceless. Furthermore, we require that the basis operators are trace-orthogonal:
\begin{equation}
{\rm Tr}\big[ \hat X_\alpha^{i'}\hat X_\beta^{j'}\big]=D\delta_{\alpha\beta}\delta_{{i'}{j'}}.
\label{eq:operator_basis_norm}
\end{equation}
 In this way any operator $\hat O^{i'}$ within a cluster $i'$ can be written as a linear combination of the basis operators:
\begin{equation}
\hat O^{i'}=\sum_\alpha o_\alpha \hat X_\alpha^{i'}.
\end{equation}

Note that the choice of clusters is not unique. Moreover a particular cluster choice can break some underlying symmetries like translational or (discrete) rotational invariance. In this case one should simply average the results over different clustering choices, which effectively restores these symmetries.

\item Define structure constants $f_{\alpha\beta\gamma}$:
\begin{eqnarray}
[\hat X_\alpha^{i'},\hat X_\beta^{j'}]&=&i f_{\alpha\beta\gamma}\delta^{{i'}{j'}} \hat X_{\gamma}^{i'}\nonumber\\ \Leftrightarrow\;
i f_{\alpha\beta\gamma}&=&{1\over D}{\rm Tr}\big[\hat X_\gamma^{i'}[\hat X_\alpha^{i'},\hat X_\beta^{i'}]\big].
\label{eq:structure_constants}
\end{eqnarray}
These two definitions are obviously equivalent. The second definition has an advantage that it can even be applied to basis operators which do not form a complete basis. Then structure constants define projectors of the commutators to the basis spanned by $\{\hat X_\alpha^{i'}\}$.

\item Rewrite the Hamiltonian $\hat H$ in terms of cluster operators
\begin{equation}
  \hat H = \sum_{{i'}{j'}} \tilde{J}_{\alpha\beta}^{{i'}{j'}}\hat X_\alpha^{{i'}}\hat X_\beta^{j'} + \sum_{i'} \tilde{B}_\alpha^{{i'}} \hat X_\alpha^{{i'}}.
  \label{eq:ham_clusterops}
\end{equation}
The new couplings $\tilde J$ and fields $\tilde B$ may be different then the base couplings; for example, local fields now include couplings between spins local to a cluster, as an operator $\hat \sigma_a^{(i)}\hat \sigma_b^{(j)}$ is linear in $\hat X_\alpha^{i'}$ if both $(i)$ and $(j)$ belong to the same cluster $i'$.

\item Identify basis operators $\hat X_\alpha^{i'}$ to classical phase space variables $x_\alpha^{i'}$ satisfying canonical Poisson bracket relations defined through the structure constants $f_{\alpha\beta\gamma}$:

\begin{equation}
\label{eq:Bopp_rep}
\hat X_\alpha^{i'} \to x_\alpha^{i'} - {i\over 2}x_\beta^{i'} f_{\alpha\beta\gamma}{\partial\over \partial x_\gamma^{i'}};\quad \{x_\alpha^{i'},x_\beta^{j'}\}=-\delta_{i'j'} f_{\alpha\beta\gamma} x_\gamma^{i'}.
\end{equation}
These Poisson brackets are obtained from a standard rule $i[\hat A,\hat B]\to \{A(\{x\}),B(\{x\})\}$, where $A,B$ are functions in on phase space variables $\{x\}$.

\item Using the operator identification, define the Hamiltonian and any observables of interest in as functions of the classical phase space variables $\{x\}$, E.g. for an observable local to a particular cluster $i'$
\begin{eqnarray}
 \hat O = \sum_{\alpha}o_\alpha \hat X_\alpha^{i'} &\to& O_W(\{x\})=\sum_{\alpha}o_\alpha x_\alpha^{i'};\quad o_\alpha={1\over D}\text{Tr}[\hat O\hat X_\alpha^{(i')}],\nonumber\\
 \hat H &\to&H_W(\{x\}) = \sum_{{i'}{j'}} \tilde{J}_{\alpha\beta}^{{i'}{j'}} x_\alpha^{{i'}}x_\beta^{j'} + \sum_{i'} \tilde{B}_\alpha^{{i'}} x_\alpha^{{i'}},
\end{eqnarray}
where index $W$ indicates that this is the Weyl symbol corresponding to the symmetric ordering. For operators linear in the cluster variables Weyl ordering does not play a role, but for operators nonlinear in $\hat X_\alpha^{(i')}$, which e.g. show up in various response functions it is important. Correct ordering always follows if one uses the Bopp representation~\eqref{eq:Bopp_rep}.

\item Define an approximate Gaussian Wigner function $W(\{x\})$ describing initial conditions.
For simplicity we only explicitly consider here pure initial states which factorize between clusters such that $W(\{x\})=\prod_{i'} W_{i'}(\{x^{i'}_\alpha\})$
\begin{equation}
W_{i'}(\{x\})={1\over Z}\exp\left[(x_\alpha-\rho_\alpha^{i'}) \Sigma^{{i'}}_{\alpha\beta} (x_\beta-\rho_\beta^{i'}) \right].
\label{eq:gaussian_wigner_function}
\end{equation}
Determine the coefficients $\rho_\alpha^{i'}$ and $\Sigma_{\alpha\beta}^{i'}$ using the initial density matrix on the ${i'}$th cluster $\hat \rho^{i'}$ by fixing the expectation values and fluctuations of the basis operators:
\begin{eqnarray}
&&{\rm Tr}\big[\hat \rho^{i'} \hat X_\beta^{i'}\big]=\int \prod_\alpha dx_\alpha\, x_\beta\, W_{i'}(\{x\}),\nonumber\\
&&{\rm Tr}\big[\hat \rho^{i'} (\hat X_\beta^{i'}\hat X_\gamma^{i'}+\hat X_\gamma^{i'}\hat X_\beta^{i'})\big]=2\int \prod_\alpha dx_\alpha\, x_\beta x_\gamma W_{i'}(\{x\})
\label{eq:gaussian_sampling}
\end{eqnarray}

Further discussion on this choice of Wigner function is given in section \ref{sec:gauss_wigner_function_details}; fixing the terms in this way corresponds to matching quantum expectation values and fluctuations with their associated values in phase space. This choice guarantees asymptotic accuracy of CTWA at short times.

\item Solve the classical equations of motion for phase space variables:
\begin{equation}
 {dx^{i'}_\alpha(t)\over dt}=-\{x_\alpha^{i'},H_W\}=f_{\alpha\beta\gamma}{\partial H_W\over \partial x_\beta^{i'}}x_\gamma^{i'}.
\label{eq:twa_eq0}
\end{equation}
Initial conditions $x_\alpha^{i'}(t=0)$ are randomly sampled from the Gaussian Wigner function and independently evolved. Note that these equations are identical to the Dirac mean-field equations if we assume that the density matrix is always factorized. Formally mean-field is recovered from the cluster TWA by setting the inverse of the fluctuation matrix $\Sigma_{\alpha\beta}$ in Eq.~\eqref{eq:gaussian_wigner_function} to zero. 

\item To find an expectation value of some observable $\hat O(t)$ in Heisenberg representation at some time $t>0$, average the corresponding classical function for each point $\{x(t)\}_n$ in phase space, where $n$ is the $n$-th sample from the Wigner function. 

\begin{equation}
\langle \hat O(t)\rangle=\lim_{N\to\infty}{1\over N}\sum_{n=1}^N O_W(\{x(t)\}_n)=\overline{O_W(\{x(t)\})}\label{eq:operator_observables}
\end{equation}

The over-line represents averaging with respect to the initial conditions sampled by the Gaussian Wigner function.

\item If one is interested in the non-equal time correlation functions then instead of the previous step use
\begin{eqnarray}
&&\langle \hat X_\alpha^{i'} (t_1) \hat X_\beta^{j'}(t_2)+  \hat X_\beta^{j'}(t_2) \hat X_\alpha^{i'} (t_1)\rangle\to 2\overline {x_\alpha^{i'}(t_1)x_\beta^{j'}(t_2)}\nonumber\\
&&i\langle [\hat X_\alpha^{i'} (t_1), \hat X_\beta^{j'}(t_2)]\rangle=f_{\alpha\gamma\nu}\, \overline {x_\gamma^{i'}(t_1) {\partial x_\beta^{j'}(t_2)\over \partial x_\nu^{i'} (t_1)}}.\label{eq:nonequaltime_correlators}
\end{eqnarray}
Here, the derivative stands for the response of $x_\beta^{j'}$ at time $t_2$ due to an infinitesimal perturbation of  $x_\nu^{i'} (t_1)$ at time $t_1$: $x_\nu^{i'} (t_1)\to x_\nu^{i'} (t_1)+\epsilon$. To preserve causality we assume that $t_1\leq t_2$, otherwise one should swap the operators $\hat X_\alpha^{i'} (t_1)$ and $\hat X_\beta^{j'}(t_2)$ and the corresponding phase space variables (see Ref.~\onlinecite{Polkovnikov2010} for details).  Using the Bopp operators, one may compute out of time order correlators in this fashion as well.

\end{enumerate}

\subsection{Wave function (Schwinger boson) CTWA}\label{sec:SB_TWA}

Instead of using $D^2$ phase space variables per cluster $x_\alpha$, one can use the Schwinger boson representation of the operators $\hat X_\alpha$ and the associated phase space, which has dimensionality $D$. In section~\ref{sec:dim_red_CTWA} we show how a similar dimensional reduction can also be done for the operator CTWA. The essential steps for the associated TWA are very similar except that we operate with standard bosonic field Poisson brackets. Instead of the Schwinger boson representation one can use e.g. an angular momentum or any other representation of the basis operators. 

\begin{enumerate}

\item Define the basis operators $\hat X_\alpha^{i'}$ in the same way as in the operator TWA (see item 1. in Sec.~\ref{sec:operator_twa}).

\item Define Schwinger boson representation of the basis operators:
\begin{equation}
\hat X_\alpha^{i'}=\sum_{a,b=0}^{D-1} \hat b_{a,{i'}}^\dagger T_\alpha^{ab} \hat b_{b,{i'}},
\end{equation}
where the matrices $\{T_\alpha\}$ form the fundamental representation of the algebra spanned by the operators $\hat X_\alpha^{i'}$. Here $\hat b^\dagger_{b,{i'}}$ is the raising operator for cluster ${i'}$ and site $b$, corresponding to the $b$-th basis state $|b\rangle$. A simple way to define these matrices is through first fixing some basis within each cluster like
\begin{eqnarray}
\label{eq:cluster_basis}
|0\rangle&=&|\uparrow,\uparrow,\dots \uparrow\rangle\\\nonumber |1\rangle&=&|\downarrow,\uparrow,\dots \uparrow\rangle\\\nonumber |D-1\rangle&=&|\downarrow,\downarrow,\dots \downarrow\rangle
\end{eqnarray}
and then defining
\[
T_\alpha^{ab}=\langle a| \hat X_\alpha |b\rangle.
\]

\item Represent the Hamiltonian (and all other observables of interest) through the Schwinger bosons

\begin{equation}
\hat H = \sum_{{i'}{j'}} \tilde J^{{i'}{j'}}_{\alpha\beta}T^{ab}_\alpha T^{cd}_\beta \hat b_{a,{i'}}^\dagger \hat b_{b,{i'}} \hat b_{c,{j'}}^\dagger \hat b_{d,{j'}} + \sum_{{i'}}\tilde B^{i'}_\alpha T_\alpha^{ab}\hat b_{a,{i'}}^\dagger \hat b_{b,{i'}} .
\end{equation}
\item Associate Schwinger boson creation and annihilation operators with complex phase space amplitudes $\hat b_{a,{i'}}^\dagger\to b_{a,{i'}}^\ast$ and $\hat b_{a,{i'}}\to b_{a,{i'}}$ satisfying canonical Poisson bracket relations:
\begin{equation}
\{b_{a,{i'}}^\ast, b_{b,{j'}}\}=-i \delta_{ab}\delta_{{i'},{j'}}.
\end{equation}
 In this way, operators get mapped to functions in phase space $\hat O\to O_W(\vec b,\vec b^*)$. For example the Hamiltonian \eqref{eq:hamiltonian} reads
\begin{multline}
H(\vec b,\vec b^*)= \\
= \sum_{{i'}{j'}} J^{{i'}{j'}}_{\alpha\beta}T^{ab}_\alpha T^{cd}_\beta b_{a,{i'}}^* b_{b,{i'}} b_{c,{j'}}^* b_{d,{j'}} + \sum_{{i'}}B^{i'}_\alpha T_\alpha^{ab}b_{a,{i'}}^* b_{b,{i'}}.
\end{multline}

\item Define an approximate Gaussian Wigner function in terms of Schwinger bosons. For the pure state $|0\rangle$ the Wigner function reads

\begin{multline}\label{eq:SB_wignerfunction}
W(\vec b,\vec b^\ast)=\delta( |b_0|^2-1-r_D)\prod_{a=1}^{D-1} \mathrm e^{-{b_a^\ast b_a\over r_D}},\\
 r_D={\sqrt{1+D}-1\over D}.
\end{multline}
Note that this Wigner function is not a standard Wigner function for Schwinger bosons, which would have a noise $1/2$ per each empty bosonic mode~\cite{Polkovnikov2010}. The phase space variables $b_{a,i'}$ can be interpreted as wave function amplitudes in the cluster basis $|0\rangle,\dots |D-1\rangle$, for a wave function which factorizes between clusters. For any other pure state related to $|0\rangle$ by some unitary rotation $|\psi_0\rangle=U|0\rangle$ the associated Wigner function can be obtained from the one in Eq.~\eqref{eq:SB_wignerfunction} by applying the unitary rotation to Schwinger bosons within a cluster: $\vec b_{i'}\to U \vec b_{i'}$. Because any unitary transformation is canonical (it preserves the Poisson brackets) the Wigner function and the equations of motion are invariant under basis transformations. More details on this rotation can be found in section \ref{sec:general_ICs}. As in the operator case, formally mean-field can be obtained from TWA by setting the fluctuations in the Wigner function~\eqref{eq:SB_wignerfunction} to zero. Note that for the fluctuating initial state $\sum_a |b_{a,{i'}}|^2>1$ so the identification of $b_{a,{i'}}$ with the wave function amplitudes is only qualitative but the equations of motion are not affected by the normalization. More details are found in section \ref{sec:SB_TWA_details}.

\item Solve the classical (Gross-Pitaevskii) equations of motion for the phase space variables $\vec b$:
\begin{equation}\label{eq:SB_evolution}
i {\partial b_{a,i'}\over \partial t}={\partial H_w\over \partial b^\ast_{a,i'}},
\end{equation}
satisfying the initial conditions drawn from the initial Gaussian Wigner distribution. As in the operator TWA these equations of motion are identical to the Dirac mean-field equations of motion obtained under the assumption that the wave-function factorizes over clusters: $|\psi(t)\rangle=\prod_{i'} |\psi_{i'}(t)\rangle$. 

\item To find the expectation value of some observable $\hat O(t)$ at some time $t>0$ average the corresponding classical function for each point $(\vec b^*_n(t),\vec b_n(t))$ in phase space, where $n$ is the $n$th sample from the Wigner function:

\begin{equation}
\langle\hat O(t)\rangle=\lim_{N\to\infty}{1\over N}\sum_{n=1}^N O_W(\vec b_n(t),\vec b_n^*(t))=\overline{O_W(\vec b(t),\vec b^*(t))}.
\end{equation}

\item If one is interested in the non-equal time correlation functions of some observables $\hat O(t_1)$ and $\hat Q(t_2)$ then instead of the previous step use
\begin{eqnarray}
&&\langle \{\hat O (t_1), \hat Q(t_2)\}\rangle\to 2\overline{ O_W(\vec b(t_1),\vec b^*(t_1))Q_W(\vec b(t_2),\vec b^*(t_2)}, \\
&&i\langle [\hat O (t_1), \hat Q(t_2)]\rangle\to \overline {\{ O_W(t_1), Q_W(t_2)\}}.
\end{eqnarray}
where $\{O_W(t_1), Q_W(t_2)\}$ is the usual non-equal time Poisson bracket (c.f. Eq. (46) in Ref.~\onlinecite{Kolodrubetz2017}):
\begin{multline}
\{O_W(t_1), Q_W(t_2)\}=\\
i\sum_{a,{i'}} {\partial O_W(t_1)\over \partial b_{a,{i'}}^\ast(t_1)}{\partial Q_W(t_2)\over \partial b_{a,{i'}}(t_1)}-
{\partial O_W(t_1)\over \partial b_{a,{i'}}(t_1)}{\partial Q_W(t_2)\over \partial b_{a,{i'}}^\ast(t_1)}.
\end{multline}

\end{enumerate}

\subsection{Example: cluster TWA for four coupled spins.}\label{sec:CTWA_example}

\begin{figure}
	\includegraphics[width=0.7\linewidth]{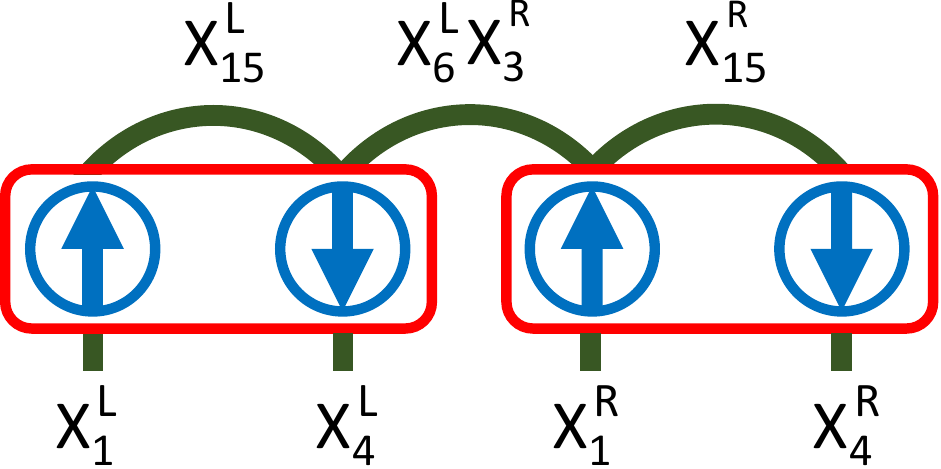}
	\caption{\textbf{Representation of the 4-spin Ising example}. Blue circles and arrows represent the physical spins in the initial Neel state, while red rectangles show the clusters labeled by $i'=L,R$. Green arches represent the nearest-neighbor $\hat \sigma_z^{(j)}\hat\sigma_z^{(j+1)}$ couplings, with the relevant representation through the cluster operators $\hat X_{\alpha}^{L,R}$. Bottom lines represent the on-site $\hat \sigma_x^{(j)}$ fields in terms of the cluster operators. In this way the original Hamiltonian~\eqref{eq:hamiltonian_four_sites} exactly maps to the two-cluster amiltonian~\eqref{eq:two_site_hamiltonian_X}}\label{fig:4spin_picture}
\end{figure}

Let us illustrate how one can apply cluster TWA to a simple system of four coupled spins shown in Fig.~\ref{fig:4spin_picture}, which is described by the following Hamiltonian:
\begin{equation}
\hat H=J\sum_{j=1}^3 \hat \sigma_z^{(j)}\hat \sigma_z^{(j+1)} +h_x\sum_{j=1}^4 \hat \sigma_x^{(j)}.
\label{eq:hamiltonian_four_sites}
\end{equation}
In this example one can consider clusters of size one, two, and four. The size one cluster leads to standard TWA for spins~\cite{Polkovnikov2010}, while the size four cluster leads to exact representation of dynamics by TWA. Size two clusters should be still approximate but lead to an improved accuracy of the method compared to the standard TWA, and will be detailed here, going through the steps highlighting how they are implemented in this example.

\subsubsection{Operator cluster TWA} \label{sec:operatorclusterTWA}

\begin{enumerate}

\item Our model system is split into two size two clusters, $1'\equiv L$ and $2'\equiv R$. The Hilbert space dimension of the cluster is $D=4$ and the dimension of the operator basis spanning $SU(4)$ algebra plus identity is $D^2=16$. A possible and convenient choice for the operator basis in the first cluster (and similarly in the second cluster) is:

\begin{equation}
\begin{matrix}
&\hat X_0=\hat I^{(1)}\otimes \hat I^{(2)}& & \\
&\hat X_1=\hat \sigma_x^{(1)} \otimes \hat I^{(2)},&\; \hat X_2=\hat \sigma_y^{(1)} \otimes \hat I^{(2)},&\;\hat X_3=\hat \sigma_z^{(1)} \otimes \hat I^{(2)}\nonumber\\
&\hat X_4=\hat I^{(1)} \otimes \hat \sigma_x^{(2)},& \hat X_5=\hat I^{(1)} \otimes \hat \sigma_y^{(2)},&\hat X_6=\hat I^{(1)} \otimes \hat \sigma_z^{(2)}\nonumber\\
&\hat X_7=\hat \sigma_x^{(1)} \otimes \hat \sigma_x^{(2)},& \hat X_8=\hat \sigma_x^{(1)} \otimes \hat \sigma_y^{(2)},&\hat X_9=\hat \sigma_x^{(1)} \otimes \hat \sigma_z^{(2)}\nonumber\\
&\hat X_{10}=\hat \sigma_y^{(1)} \otimes \hat \sigma_x^{(2)},&\hat X_{11}=\hat \sigma_y^{(1)} \otimes \hat \sigma_y^{(2)},&\hat X_{12}=\hat \sigma_y^{(1)} \otimes \hat \sigma_z^{(2)}\nonumber\\
&\hat X_{13}=\hat \sigma_z^{(1)} \otimes \hat \sigma_x^{(2)},&\hat X_{14}=\hat \sigma_z^{(1)} \otimes \hat \sigma_y^{(2)},&\hat X_{15}=\hat \sigma_z^{(1)} \otimes \hat \sigma_z^{(2)},\nonumber\\
\end{matrix}
\end{equation}
where superscripts $1$ and $2$ refer to the first and the second site of the cluster. It is easy to check that these basis operators satisfy the required normalization conditions~\eqref{eq:operator_basis_norm}.

\item The structure constants are easy to get from the commutation relations of the Pauli matrices. There are in total $20$ independent structure constants in this operator basis. Let us explicitly show a few of them:
\begin{eqnarray}
[\hat X_1,\hat X_2]&=&2i \hat X_3\;\Rightarrow f_{1,2,3}=2,\\\nonumber
[\hat X_7,\hat X_{10}]&=&2i \hat X_3\;\Rightarrow f_{7,10,3}=2.
\end{eqnarray}

\item In terms of the cluster operators the Hamiltonian~\eqref{eq:hamiltonian_four_sites} reads
  \begin{eqnarray}
\label{eq:two_site_hamiltonian_X}
  \hat H = &J& \left(\hat X_{15}^{L}+\hat X_{15}^{R}+\hat X_6^{L} \hat X_3^{R}\right)\\\nonumber
  +&h_x& \left(\hat X_1^{L}+\hat X_4^{L}+\hat X_1^{R}+\hat X_4^{R}\right).
 \end{eqnarray}
Note that this Hamiltonian contains only a single nonlinear coupling between the operators $\hat X_6^{L}$ and $\hat X_3^{R}$. When written in terms of Pauli matrices (single site clusters) the Hamiltonian has three non-linear couplings.

\item Identify fifteen phase space variables per each cluster $x_{1}^{L,R}\dots x_{15}^{L,R}$ satisfying the Poisson bracket relations with the structure constants determined above.

\item Define the classical Hamiltonian $H_W$:
\begin{eqnarray}
H_W &=&J \left(x_{15}^{L}+x_{15}^{R}+x_6^{L} x_3^{R}\right)\\&+&h_x \left(x_1^{L}+x_4^{L}+x_1^{R}+x_4^{R}\right).
\end{eqnarray}

\item Let us consider an initial product state 
\[
|\psi_0\rangle=|\uparrow,\downarrow,\uparrow,\downarrow\rangle,
\] 
which is an eigenstate of the Hamiltonian with $h_x=0$. Then our setup can be regarded as a quench from this initial state. The Gaussian Wigner function for this state factorizes into a product of left and right Wigner functions $W_L$ and $W_R$. For the left cluster (and similar for the right cluster) the nonzero expectation values have only the following three operators:
\begin{equation}
\langle \psi_0| \hat X_3^L |\psi_0\rangle=1,\; \langle \psi_0| \hat X_6^L |\psi_0\rangle=-1,\;
\langle \psi_0| \hat X_{15}^L |\psi_0\rangle=-1.
\end{equation}

There are 24 non-zero symmetric correlation functions excluding the identity for the left cluster (and similar for the right cluster):

\begin{eqnarray}
&&\left< \psi_0\big|\left(\hat X_{\alpha}^{L}\right)^2\big |\psi_0\right>=1,\; \quad\alpha\in\{1,\dots, 15\}\\
&&{1\over 2}\langle \psi_0|\{ \hat X_{\alpha}^{L}, \hat X_{\beta}^{L}\}_+ |\psi_0\rangle=1\nonumber\\
&&\qquad(\alpha,\beta)\in
\begin{bmatrix}
(4,13) & (5,14) & (6,15) & (7,11)
\end{bmatrix}\nonumber\\
&& {1\over 2}\langle \psi_0 |\{\hat X_{\alpha}^L, \hat X_{\beta}^L\}_+|\psi_0\rangle=-1\nonumber\\
&&\qquad(\alpha,\beta)\in
\begin{bmatrix}
(1,9) & (2,12) & (3,6) & (3,15) & (8,10)
\end{bmatrix},\nonumber
\end{eqnarray}	
where $\{\hat A,\hat B\}_+$ stands for anti-commutator. The fact that
\[
\langle \psi_0|\hat X_8^2|\psi_0\rangle=\langle \psi_0|\hat X_{10}^2|\psi_0\rangle=-{1\over 2}\langle \psi_0|\{\hat X_8,\hat X_{10}\}_+|\psi_0\rangle=1
\]
implies that the phase space variables $x_8$ and $x_{10}$ should be perfectly anti-correlated. After finding the connected correlators, the associated Gaussian Wigner functions $W_L$ and $W_R$ read
\begin{multline}
W_L={1\over Z}\delta(x_3-1)\delta(x_6+1)\delta(x_{15}+1)\delta(x_4-x_{13})\delta(x_5-x_{14})\\
\times\delta(x_1+x_9)\delta(x_2+x_{12})\delta(x_7-x_{11})\delta(x_8+x_{10})\\
\times\exp\left[-{\sum_{\alpha\in\{1,2,4,5,7,8\}} x_\alpha^2\over 2}\right],
\label{eq:wigner_example_L}
\end{multline}
where $Z$ is the normalization constant. The $\delta$-functions can be understood as limiting cases of Gaussians and thus are allowed in our ansatz. We see that each initial condition must draw 12 uncorrelated numbers (6 per each cluster).

\item Now we need to solve the system of classical equations of motion~\eqref{eq:twa_eq0} for our phase space variables with the initial conditions $x_\alpha^{L,R}(t=0)$ randomly sampled from the left and right Wigner functions~\eqref{eq:wigner_example_L}. There are overall thirty equations: fifteen for left variables and fifteen for right variables (the variables $x_0^{L,R}$ corresponding to identity operators are obviously conserved in time). Let us explicitly show a few of them:

\begin{eqnarray}
\partial_t x_{4}^L&=& Jx_{3}^Rx^L_{5} +Jx^L_{14}\\
\partial_t x_{3}^L&=& -hx^L_{2}\nonumber\\
\partial_t x_{12}^L&=& -hx^L_{11} +hx^L_{15} -Jx^L_{1}.\nonumber
\end{eqnarray}


\item We can now compute the expectation values of observables by averaging the corresponding Weyl symbols computed on time-dependent phase space points over the initial conditions. For example:
\begin{eqnarray}
\langle  \hat \sigma_z^{(1)}(t)\hat \sigma_z^{(2)}(t)\rangle=\langle  \hat X_{15}^L(t)\rangle&\approx& \overline{x_{15}^L(t)},\\
\langle \hat \sigma_z^{(2)}(t)\hat \sigma_z^{(3)}(t)\rangle=\langle \hat X_{6}^L(t) \hat X_3^R(t)\rangle&\approx& \overline{x_6^L(t) x_3^R(t)},\nonumber\\
\langle \hat \sigma_z^{(1)}(t) \rangle = \langle \hat X_{3}^L(t)\rangle&\approx& \overline{x_3^L(t)}.\nonumber
\end{eqnarray}

\item In order to compute a non-equal time correlation function, for example of the z-magnetization on the first site, we can use~Eq.~\eqref{eq:nonequaltime_correlators} identifying $\hat \sigma_z^{(1)}\to x_3^L$:
\begin{eqnarray}
\langle \{\hat{\sigma}_z^{(1)}(t),\hat \sigma_z^{(1)}(0)\}_+\rangle&=& 2\overline{x_3^L(t)x_3^L(0)}\\
i\langle[\hat{\sigma}_z^{(1)}(0),\hat \sigma_z^{(1)}(t)]\rangle&=&f_{3\mu\nu}\overline{\bigg(x_{\mu}^L(0){\partial x_3^L(t)\over \partial x_\nu^L(0)}\bigg)}.\nonumber
\end{eqnarray}


\end{enumerate}

\begin{figure}
	\includegraphics[width=\linewidth]{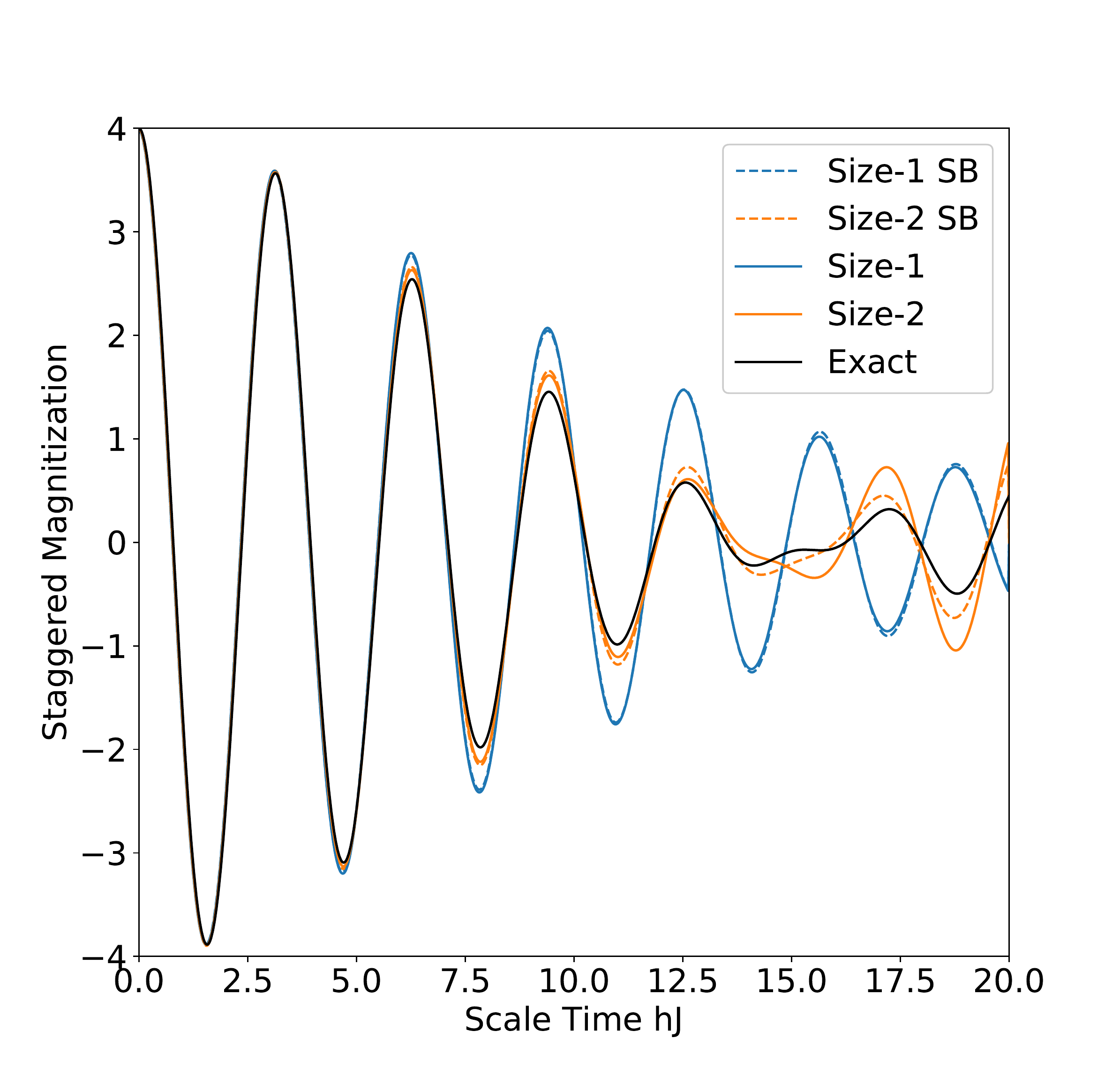}
	\caption{\textbf{Dynamics of the staggered magnetization in the 4-spin Transverse Ising example} (see text for details). Solid black represents the exact result, while the solid (dashed) colored lines are the results of the Operator (Schwinger Boson) CTWA. The staggered magnetization is represented by $\overline{x_{3}^L + x_{3}^R - x_{6}^L - x_{6}^R}$. CTWA dynamics approach exact results with  increasing cluster size.}\label{fig:IsingExample}
\end{figure}

In Fig. ~\ref{fig:IsingExample} we show the results of simulations using the cluster TWA. For comparison we also show the single cluster (conventional) TWA and exact results. For simulations we choose specific parameters: $J=\frac{1}{8},\;h_x=1$.

\subsubsection{Wave function CTWA} 
This same example system can be done under wavefunction (Schwinger Boson) CTWA.
\begin{enumerate}
	\item First, define the 15+1 basis operators for each of the two clusters in the same fashion as the Operator cluster TWA example. Define these operators as 15 4x4 matrices $\hat X_\alpha\to T^{ab}_\alpha \hat a^\dagger_a \hat a_b$ with complex coordinates for (L/R) of ($(a,a^*)$ / $(b,b^*)$). For example, the operators:
	
	\begin{eqnarray}
	\hat X_4^L=\hat I^{(1)}\otimes\hat \sigma_x^{(2)}&\to&
	\begin{bmatrix}
	\hat a_0^\dagger \\ \hat a_1^\dagger \\ \hat a_2^\dagger \\ \hat a_3^\dagger
	\end{bmatrix}^T
	\begin{bmatrix}
	0&  1&  0&  0\\
	1&  0&  0&  0\\
	0&  0&  0&  1\\
	0&  0&  1&  0
	\end{bmatrix}
	\begin{bmatrix}
	\hat a_0\\ \hat a_1\\ \hat a_2\\ \hat a_3
	\end{bmatrix},\\
	\hat X_8^L=\hat \sigma_x^{(1)}\otimes\hat \sigma_y^{(2)}&\to&
	\begin{bmatrix}
	\hat a_0^\dagger \\ \hat a_1^\dagger \\ \hat a_2^\dagger \\ \hat a_3^\dagger
	\end{bmatrix}^T
	\begin{bmatrix}
	0&  0&  0&  -i\\
	0& 0&  i&  0\\
	0&  -i& 0&  0\\
	i&  0&  0& 0
	\end{bmatrix}
	\begin{bmatrix}
	\hat a_0\\ \hat a_1\\ \hat a_2\\ \hat a_3
	\end{bmatrix}.\nonumber
	\end{eqnarray}
	
	The classical Hamiltonian $H_W$ becomes:
	
	\begin{multline}
	H_W(a,b) = (a_i^* a_j + b_i^*b_j)\big(JT^{ij}_{15} + h_xT^{ij}_3 + h_xT^{ij}_4\big)\\
	+ J\,a_i^* a_j b_k^* b_l\,T^{ij}_6 T^{kl}_3.
	\end{multline}
	
	\item Draw initial conditions for $(\vec a,\vec b)$. For the state $|\uparrow\downarrow\uparrow\downarrow\rangle$ the initial condition is $|\uparrow\downarrow\rangle$ per cluster corresponds to $a_1,b_1 = \sqrt{1 + \delta_0}$ and $a_q,b_q$ with $q=0,2,3$ random numbers drawn from the Gaussian probability distribution~\eqref{eq:SB_wignerfunction} with zero mean and the variance $\delta_0={\sqrt{1+2^2}-1\over 2^2}\approx 0.30902$. Note that because the initial state here is $\uparrow\downarrow\rangle$ and not $|\uparrow\uparrow\rangle$, it is the Schwinger bosons $a_1,\,b_1$ (not $a_0,\, b_0$), which are special. Similarly for $|\downarrow\uparrow\rangle$ ($|\downarrow\downarrow\rangle$) states the bosons $a_2,\,b_2$ ($a_3,\,b_3$) will be special.
	
	\item Evolve each point $(\vec a(t),\vec b(t))$ via Gross-Pitaevskii equations~\eqref{eq:SB_evolution}, which explicitly read:
	\begin{eqnarray}
	&&i{\partial\over\partial t}
	\begin{bmatrix}
		a_0\\a_1\\a_2\\a_3
	\end{bmatrix}
	=
	\begin{bmatrix}
	J & h & h & 0\\
	h & -J & 0 & h\\
	h & 0 & -J & h\\
	0 & h & h & J
	\end{bmatrix}
	\begin{bmatrix}
	a_0\\a_1\\a_2\\a_3
	\end{bmatrix}\nonumber\\
	&&+ J(|b_0|^2 + |b_1|^2 - |b_2|^2 - |b_3|^2)
	\begin{bmatrix}
	a_0\\-a_1\\a_2\\-a_3
	\end{bmatrix}
	\end{eqnarray}
and similarly for the $b$-bosons.

	\item We can compute expectation values of observables by averaging the corresponding Weyl symbols computed on time-dependent phase space points over the initial conditions. For example:
	
	\begin{eqnarray}
		\langle  \hat \sigma_z^{(1)}(t)\hat \sigma_z^{(2)}(t)\rangle&\approx& \overline{a_a^* a_b} T^{ab}_{15},\\
		\langle \hat \sigma_z^{(2)}(t)\hat \sigma_z^{(3)}(t)\rangle&\approx& \overline{a_a^* a_b b_x^* b_y}T_{6}^{ab}T_3^{xy},\nonumber\\
		\langle \hat \sigma_z^{(1)}(t) \rangle &\approx& \overline{a_a^* a_b}T^{ab}_3.\nonumber
	\end{eqnarray}
	
	The results of the simulations of the Schwinger boson CTWA are shown in Fig.~\eqref{fig:IsingExample} using dashed lines. They are almost indistinguishable from the operator CTWA results.
\end{enumerate}

\subsection{CTWA and the Dirac Mean Field Approximation.}


The basic idea of the CTWA (as of any other TWA) is best summarized in Fig. \ref{fig:phasespace_example}. One first generates an initial condition drawn from a Gaussian distribution~\eqref{eq:gaussian_wigner_function} or ~\eqref{eq:SB_wignerfunction}. Then the phase space variables evolve in time according to the semiclassical (or equivalently mean field) equations of motion \eqref{eq:twa_eq0} or \eqref{eq:SB_evolution}. After that, one evaluates observables of interest on these time evolved trajectories and repeats the procedure; effectively averaging over initial conditions. Note that even though classical and mean field equations are identical, TWA is not equivalent to the mean field approximation and generally outperforms mean field in non-linear (interacting) systems. TWA reduces to the mean-field approximation only if the initial Winger function contracts to a point in phase space and thereby suppress the initial quantum noise to zero.

The reason for such nonequivalence between TWA and mean-field approaches in non-linear systems is that averaging over initial conditions and time evolution are non-commuting operations. Therefore time evolving an average trajectory is generally very different from first solving nonlinear equations of motion for randomly chosen initial conditions and then averaging the observable of over the initial noise. In fact using short time perturbation theory one can show that CTWA correctly reproduces exact dynamics of observables up to the order $t^2$, while the mean field dynamics is accurate only up to the order $t$ (For more details, see section \ref{sec:gauss_wigner_function_details}). This observation also justifies sufficiency of the Gaussian initial conditions because even if we improve further the Wigner function matching the higher initial moments of phase point operators, the accuracy of CTWA will remain the same because there will be corrections of the order of $t^3$ coming from quantum jumps, which are beyond TWA~\cite{Polkovnikov2010}.

In situations where one has large statistical fluctuations, e.g. the initial density matrix represents an infinite temperature state like it was done in Ref.~\onlinecite{Leviatan2017} the statistical noise dominates over the quantum noise. This is reflected in the Wigner function of the infinite temperature density matrix being essentially completely random. Even in those cases, we believe, referring to the method proposed by the authors as variational mean field approximation is somewhat misleading, because each initial condition is evolved independently and the averaging is done in the end. Mean field dynamics, by construction implies that a single initial condition represents the dynamics of the whole system.

CTWA is also similar in spirit to the variational approach proposed in Ref.~\onlinecite{Shi2017} if we assume that the Wigner function remains Gaussian at all time. Then instead of averaging over initial conditions one simply has to propagate both average and variance of phase space variables. We comment, however, that this would introduce another approximation into the CTWA method, because for a nonlinear evolution any initial Gaussian distribution function becomes non-Gaussian. In some situations, for example where classical dynamics becomes unstable, the Gaussian approximation can become very inaccurate. A simple example where their approximation will fail would be dynamics of a particle initialized on the top of a Mexican hat potential (see Refs.~\onlinecite{Polkovnikov2010, Sels2012}). In other situations the Gaussian approximation for the Wigner function might remain accurate even after long times.

  \section{Application of CTWA to the non-integrable Ising Model}\label{sec:Ising_results}
	
	As a first nontrivial example, we present the method as applied to the Ising model in a transverse and longitudinal field, which is known to have a quantum chaotic regime. Here, we consider the setup similar to that of Ref.~\onlinecite{Leviatan2017} where we initially polarize a single spin in the up state and follow time decay of the magnetization. However, unlike Ref.~\onlinecite{Leviatan2017} we consider two setups of the ``remaining'' bath spins where i) they are prepared in a pure polarized state and ii) they are prepared in a mixed state. We consider the Hamiltonian:
\begin{equation}
\label{eq:Hamiltonian_spin_diffusion}
	\hat{H}=\sum_{i=0}^{19} \hat \sigma_x^{(i)} \hat \sigma_x^{(i+1)} +0.8090\hat \sigma_x^{(i)} - 0.9045\hat \sigma_z^{(i)},
	\end{equation}
with periodic boundary conditions. This set of couplings is known to lead to the chaotic Hamiltonian satisfying the eigenstate thermalization hypothesis~\cite{Kim2014}. To benchmark our CTWA results we compare them to exact quantum dynamics for a system of 20 spins. For the pure state initial condition we choose $|\psi\rangle = |\uparrow\downarrow\downarrow\dots\downarrow\rangle$, i.e. all spins except the first one are polarized along the $-z$ direction. For the mixed state initial condition we uniformly sample all the bath spins from the Bloch sphere keeping always the first spin polarized along the positive $z$-axis. The latter setup is identical to the one studied in Ref.~\onlinecite{Leviatan2017}.

\begin{figure}[h]
		\includegraphics[width=0.9\linewidth]{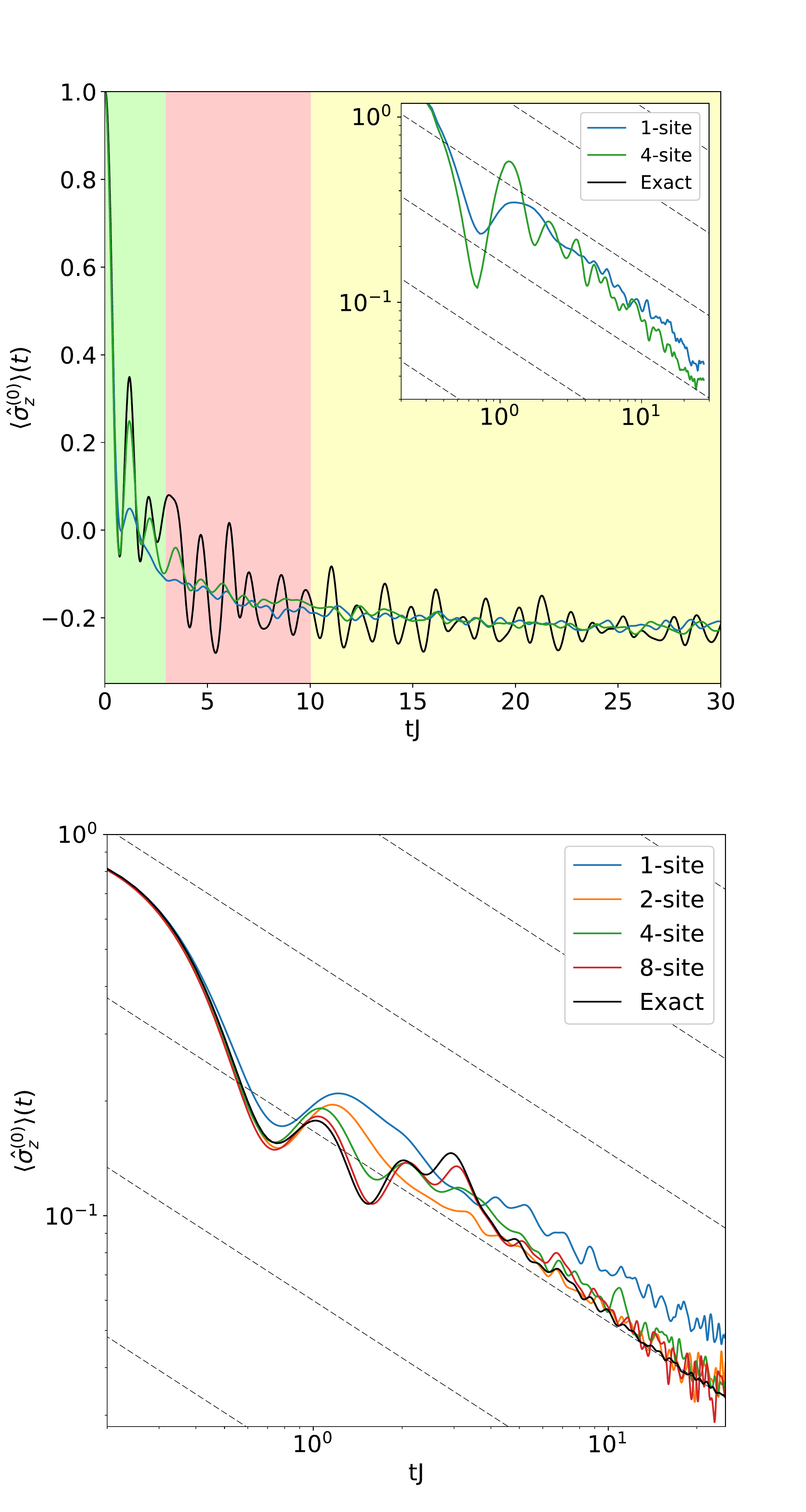}
		\caption{\textbf{Magnetization decay in a spin chain.} Time dependence of $\langle \hat\sigma_z^{(0)}(t)\rangle$ of the initially polarized spin coupled to other spins forming a ``bath''. The top plot corresponds to the initial pure state $|\uparrow\downarrow\dots\downarrow\rangle$.The bottom plot corresponds to the mixed, infinite temperature, state of the bath, where all spins except one are prepared in a random initial state. The system size is $N=20$ for the pure state and $N=16$ for the mixed state; the Hamiltonian of the system is given by Eq.~\eqref{eq:Hamiltonian_spin_diffusion}. The inset in the top plot shows $\langle \hat \sigma_z^{(0)}(t)\rangle - {1/N}\sum_j \langle \sigma_z^{(j)}\rangle$. Both setups show initial coherent oscillations of the magnetization followed by long-time diffusive relaxation to the thermal state. The accuracy of the CTWA clearly improves with the cluster size. Dashed lines are $1/\sqrt{t}$ diffusive asymptotes. }
\label{fig:ising-decay}
	\end{figure}

In Fig.~\ref{fig:ising-decay} we show time dependence of the $z$-magnetization of the initially $z$-polarized spin for the pure state of the bath (top) and the mixed state of the bath (bottom). For the pure initial state the CTWA reproduces short-time coherent oscillations, as well  as long-time diffusive decay and thermodynamic equilibrium. The only thing CTWA misses are the intermediate time oscillations of the magnetization; this mistake decreases for larger clusters. The agreement of CTWA with exact dynamics is even better for the mixed initial state of the bath.

While the accuracy of CTWA at short times can be generally expected as in standard TWA approximations, correctly capturing the long time dynamics is highly nontrivial as TWA often leads to an uncontrolled error~\cite{Polkovnikov2010}. Intuitively this feature of CTWA is not that surprising as cluster variables correctly capture short-distance quantum correlations, while the long time/long distance dynamics is generally expected to be classical. At least for thermalizing, nonintegrable, systems CTWA is expected to predict accurately both short time and long time results with the increasing cluster size as seen in the figure (see also Ref.~\onlinecite{Davidson2017}, where similar long time accuracy of the fermionic TWA was observed).

\textbf{Short time dynamics}. Like the traditional TWA (see Ref.~\onlinecite{Polkovnikov2010} for details) CTWA is asymptotically exact at short times, up to the order $\mathcal O(t^2)$. This is guaranteed by the short time perturbation theory for all observables. Specifically any short time response of an observable $\hat O$ can be Taylor expanded in time, as is detailed in section \ref{sec:gauss_wigner_function_details}. Similarly, one may heuristically argue that dynamics are exact until entanglement between clusters, which is treated approximately, become relevant.

 \textbf{Long time dynamics.} In CTWA we generically deal with classical nonlinear and hence chaotic systems. For linear systems CTWA is guaranteed to be exact at all times and the case of nonlinear integrable systems goes beyond the scope of this paper and requires a special attention. Then the long time steady state is expected to be described by the thermal equilibrium:
\begin{equation}
\langle \hat O(t)\rangle_{t\to\infty}\to =Z^{-1}\int d\vec x e^{-\beta H_W(\{x\})}O_W(\{x\}),
\end{equation}
where $\beta$ is the temperature set by the initial energy of the system. At least for high temperatures this Gibbs distribution is equivalent to the quantum thermal distribution $\langle \hat O\rangle =\textnormal{Tr}[\hat O e^{-\beta(E)\hat H}]$. Moreover the long time approach to the thermal equilibrium, is usually described within the classical hydrodynamic framework and thus is compatible with CTWA. While we do not have more mathematically rigorous arguments of asymptotic equivalence of thermal classical (in terms of cluster variables) and quantum distributions, we observed that in all ergodic regimes as we increase the cluster size CTWA correctly predicts long time behavior of chaotic quantum systems. 
	
	\textbf{Intermediate time dynamics.} CTWA may miss intermediate time dynamics of the system, as illustrated in Fig.~\ref{fig:ising-decay}. The mistake is usually can be controlled by increasing the cluster size. For sufficiently large clusters the system might enter the classical hydrodynamic regime before the mistake due to the truncation of quantum dynamics kicks in. Then the CTWA becomes essentially exact at all times like it e.g. happens for the 8-site CTWA in the bottom plot of Fig.~\ref{fig:ising-decay}. This usually happens at high temperatures and away from integrability. Close to the ground state the coherent quantum dynamics is expected to persist for very long times and as a result one needs to use very large clusters to capture dynamics correctly at long times. This can be potentially achieved by combining CTWA with DMRG techniques and is a subject of future work.

  \section{Application of CTWA to the Disordered 1d Heisenberg Chain}\label{sec:1d_results}
  
  Another system to test CTWA is the 1d Heisenberg model with evenly distributed disorder in the $z$-component of the magnetic field and periodic boundary conditions:
	
	\begin{equation}\label{eq:heisenberg_hamiltonian}
	\hat{H}= J\sum_{\langle ij\rangle}\big(\hat{\sigma}_x^{(i)}\hat{\sigma}_x^{(j)} + \hat{\sigma}_y^{(i)}\hat{\sigma}_y^{(j)} +\hat{\sigma}_z^{(i)}\hat{\sigma}_z^{(j)}\big) + h_z\sum_{i=1}^N\delta^i\hat{\sigma}_z^{(i)},
	\end{equation}
where $\delta^i\in[-1,1]$. It is well known phenomenologically that this model undergoes a transition to the Many Body Localized (MBL) phase~\cite{Luitz2015} above some critical disorder strength $h_z\approx3.5$. This phase is characterized, amongst other things, by a "memory" of initial conditions which persists indefinitely, as well as a logarithmic growth of entanglement in time after a quench. A natural order parameter to measure the transition to MBL is the staggered $z$-magnetization $\hat M_z$~\cite{Luitz2015,Leviatan2017}, which is maximal at $t=0$ for a Neel initial state $|\psi_0\rangle$:
\begin{equation}\label{eq:neelstate}
	|\psi(0)\rangle=|\dots\uparrow\downarrow\uparrow\downarrow\uparrow\downarrow\dots\rangle\quad;\quad \hat M_z=\sum_{i=1}^N(-1)^i \hat{\sigma}^{(i)}_z.
	\end{equation}
If the system thermalizes then this order parameter decays to zero; while in MBL non-thermal regime it decays to some non-zero value. 

Because averaging over disorder and quantum fluctuations are commuting operations we can simultaneously update initial conditions according to the Wigner function and the disorder realization of the Hamiltonian. This allows one to improve numerical convergence of the results and parallelize computations if necessary.

\begin{figure*}
		\centering
		\includegraphics[width=\linewidth]{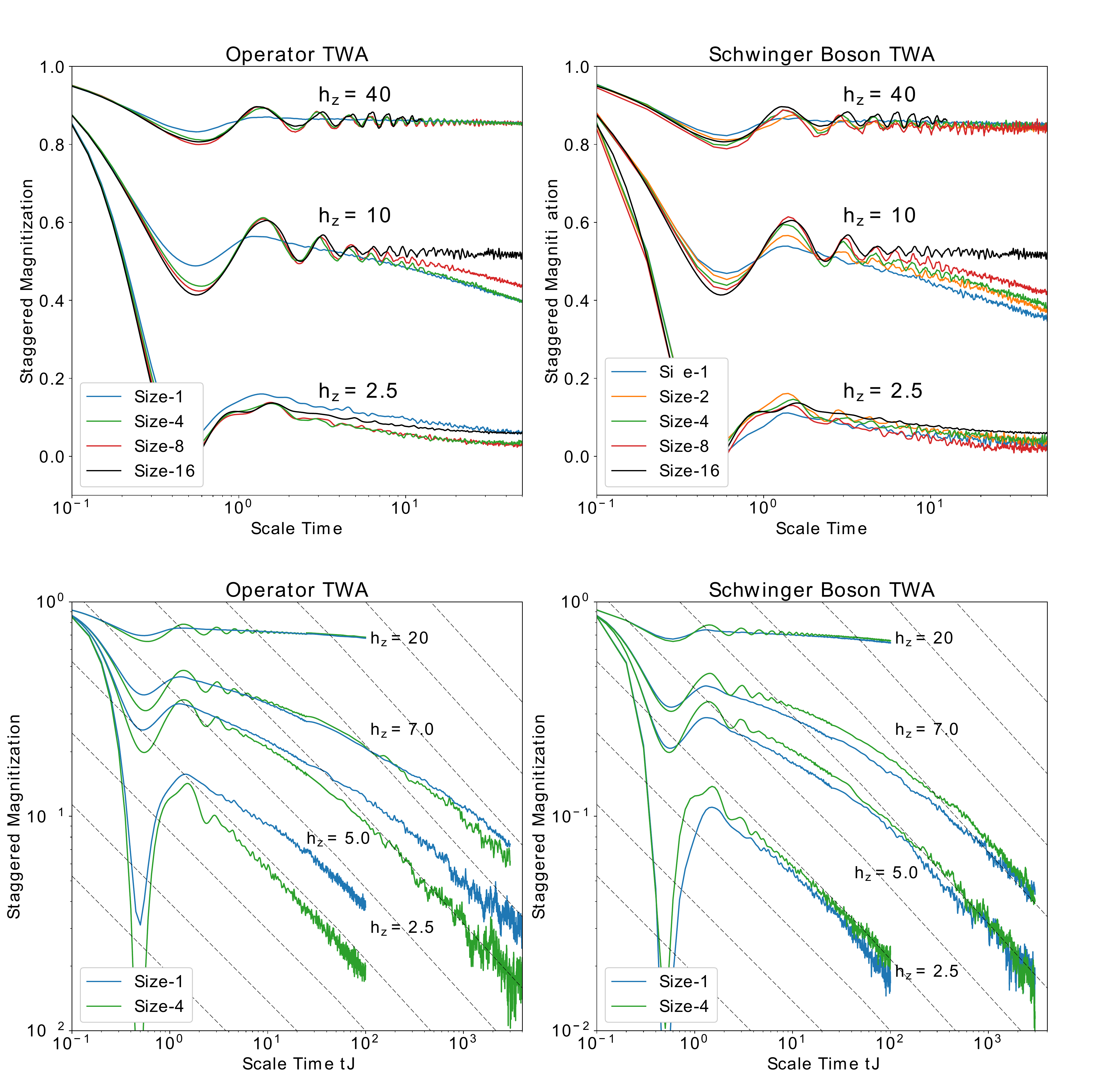}
		\caption{ \textbf{Dynamics of the staggered magnetization in the 1d Disordered Heisenberg Chain}. TOP: Comparison of exact dynamics vs operator CTWA (left) and Schwinger Boson CTWA (Right) for varying cluster sizes and a 16-site Heisenberg chain (see text for details). BOTTOM: long time decay of the staggered magnetization within the cluster CTWA (Left) and Schwinger Boson CTWA (Right) for a 64-site 1d Heisenberg chain. Straight lines show $1/\sqrt{t}$ diffusive asymptotes of the decay.}\label{fig:exactcompare_DTWA} 
	\end{figure*}
%
	
In Fig.~\ref{fig:exactcompare_DTWA} we plot the normalized staggered magnetization 
\[
m_z={1\over N}\langle \psi(t) |\hat M_z |\psi(t)\rangle,
\]
for the initial Neel state and different strengths of disorder. The left two plots show the results of the operator CTWA and the right plots show wave function (Schwinger Boson) CTWA. Both methods clearly lead to nearly identical curves. On the top two plots we show results for a relatively small system size $N=16$, which allows for comparison with exact diagonalization. The bottom plots show CTWA results for a larger system size $N=64$.

At short times, CTWA reproduces very well exact quantum dynamics. Accuracy of CTWA clearly improves with the cluster size. CTWA works well both at low disorder and at high disorder, but it clearly fails to reproduce localization at intermediate values of disorder. This is consistent with earlier work~\cite{Oganesyan2009,Acevedo2017} observing that classical disordered spin chains in thermodynamic limit eventually show ergodic diffusive behavior.  Our numerical results (bottom plots) indeed indicate that irrespective of the cluster size long time dynamics with CTWA approximation is always diffusive. Nevertheless CTWA indicates presence of MBL regime showing longer and longer localized prethermalization plateaus with the increasing cluster size.

This example illustrates limitations of CTWA to correctly capture long time non-ergodic behavior. We believe that this is not a fundamental limitation though and even in this case CTWA can be improved by choosing a better operator basis, for example using not cluster operators constructed from Pauli matrices but from so called $l$-bits~\cite{Pal2017}, which also form a complete operator basis or using the basis of the matrix product operators.~\cite{Pirvu2010}

\subsection{Entanglement entropy within CTWA}
Interestingly CTWA allows one not only to extract expectation values of local observables but also get information about entanglement of subsystems. In order to compute the entanglement entropy we need to treat the full density matrix of a subsystem as an observable. For simplicity we will only consider the density matrices which are either confined to a single cluster or span two clusters. Any density matrix confined to a cluster can be represented as
\[
\hat \rho^{i'}(t)=\sum_\alpha c_\alpha^{i'}(t) \hat X_\alpha^{i'},
\]
with some time dependent coefficient. From the orthonormality of the basis we obtain
\begin{multline*}
c_\alpha^{i'}={1\over D}{\rm Tr} [\hat \rho^{i'}(t) \hat X_\alpha^{i'}]={1\over D}{\rm Tr} [\hat \rho(t) \hat X_\alpha^{i'}]\\
={1\over D}{\rm Tr} [\hat \rho \hat X_\alpha^{i'}(t)]
={1\over D} \langle \hat X_\alpha^{i'}(t)\rangle
\approx {1\over D}\overline{x_\alpha^{i'}(t)},
\end{multline*}
where the last equality holds only within the accuracy of CTWA approximation. Therefore we can rewrite the reduced density matrix of a cluster (or similarly any sub-cluster of a larger cluster) as
\begin{equation}\label{eq:intra_rho}
\hat \rho^{i'}(t)\approx {1\over D}\sum_\alpha  \overline{x_\alpha^{i'}(t)}\hat X_\alpha^{i'}.
\end{equation}
Likewise for a reduced density spanning two clusters (or similarly two sub-clusters) we find
\begin{equation}\label{eq:inter_rho}
 \hat \rho^{{i'}{j'}} \approx {1\over D^2} \sum_{\alpha\beta}\overline{x_\alpha^{i'}(t) x_\beta^{j'}(t)} \hat X_\alpha^{i'}\otimes \hat X_\beta^{j'}.
\end{equation}
Note that generically for any interacting (nonlinear) Hamiltonian the averages $\overline{x_\alpha^{i'}(t) x_\beta^{j'}(t)}$ do not factorize so one can get entanglement between clusters even within CTWA like it was recently analyzed in Ref.~\onlinecite{Acevedo2017}. We note that to get the inter-cluster entanglement the presence of the noise is crucial because in any noiseless mean-filed type approximations the equality $\overline{x_\alpha^{i'}(t) x_\beta^{j'}(t)}=\overline{x_\alpha^{i'}(t)}\,\overline{x_\beta^{j'}(t)}$ always holds. The entaglement entropy of the particular subsystem $A$, which can include one or more clusters, is evaluated according to the standard rule: 
\begin{equation}\label{eq:entropy}
S^{i'}(t) = \textnormal{Tr}[\hat\rho_A (t) \textnormal{log}_2(\hat\rho_A(t))].
\end{equation}

	\begin{figure}
		\centering
		\includegraphics[width=\linewidth]{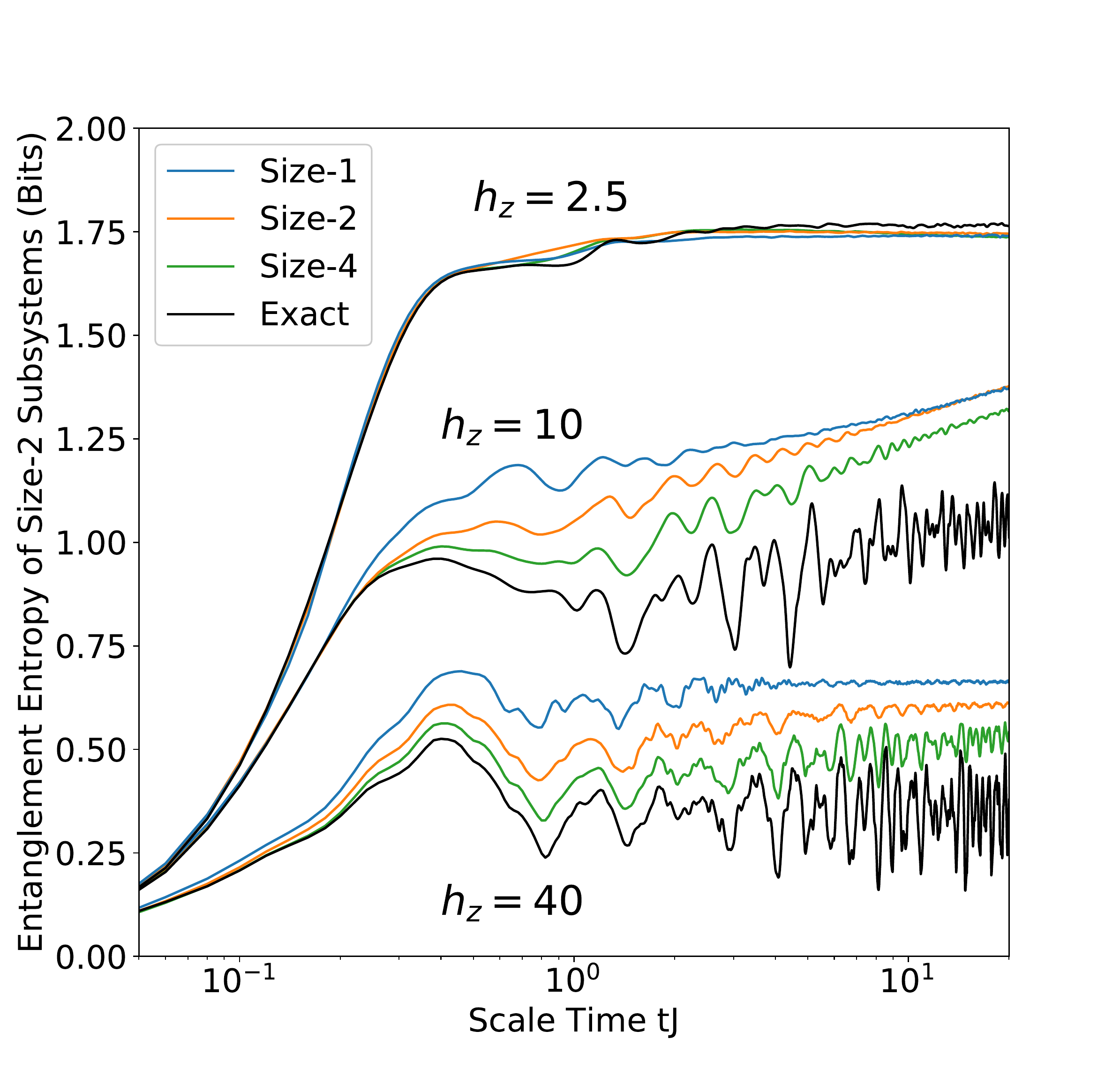}
		\caption{\textbf{Entropy Growth in the Disordered Heisenberg Model} 
		Showing the Von Neumann entropy (equation \eqref{eq:entropy}) for size-2 subsystems of a size-16 Heisenberg chain and fixed disorder realization. Entropy is averaged over all adjacent size-2 subsystems. Black is exact dynamics, colored are CTWA dynamics for various cluster sizes. All cluster offsets were used, (eg, 4 offsets for size-4 clusters). Both intra (equation \eqref{eq:intra_rho}) and inter (equation \eqref{eq:inter_rho})- cluster subsystems were averaged. Clearly the entanglement entropy approaches the exact result for larger cluster sizes.}\label{fig:3}
	\end{figure}

In Fig.~\ref{fig:3} we show the calculated entropy density for the disordered Heisenberg model of size $N=16$. At low disorder, the reduced entropy density is known to grow ballistically ($\sim t$) before saturating at a thermal value. At high disorder, i.e. in the MBL regime, the entanglement entropy shows initial growth followed by intermediate time saturation at a prethermal plateau and then eventual logarithmic growth, consistent with the results of the exact diagonalization (see the review.~\onlinecite{Altman2015} and refs. therein for further details).  As with other observables accuracy of CTWA improves with increasing cluster size. In contrast to the exact dynamics, the  slow logarithmic growth of the entanglement entropy in the CTWA goes hand in hand with decay of the order parameter. For this reason the CTWA likely overestimates the slope of asymptotic growth of the entropy with $\log(t)$.
	
	\begin{figure}
		\centering
		\includegraphics[width=\linewidth]{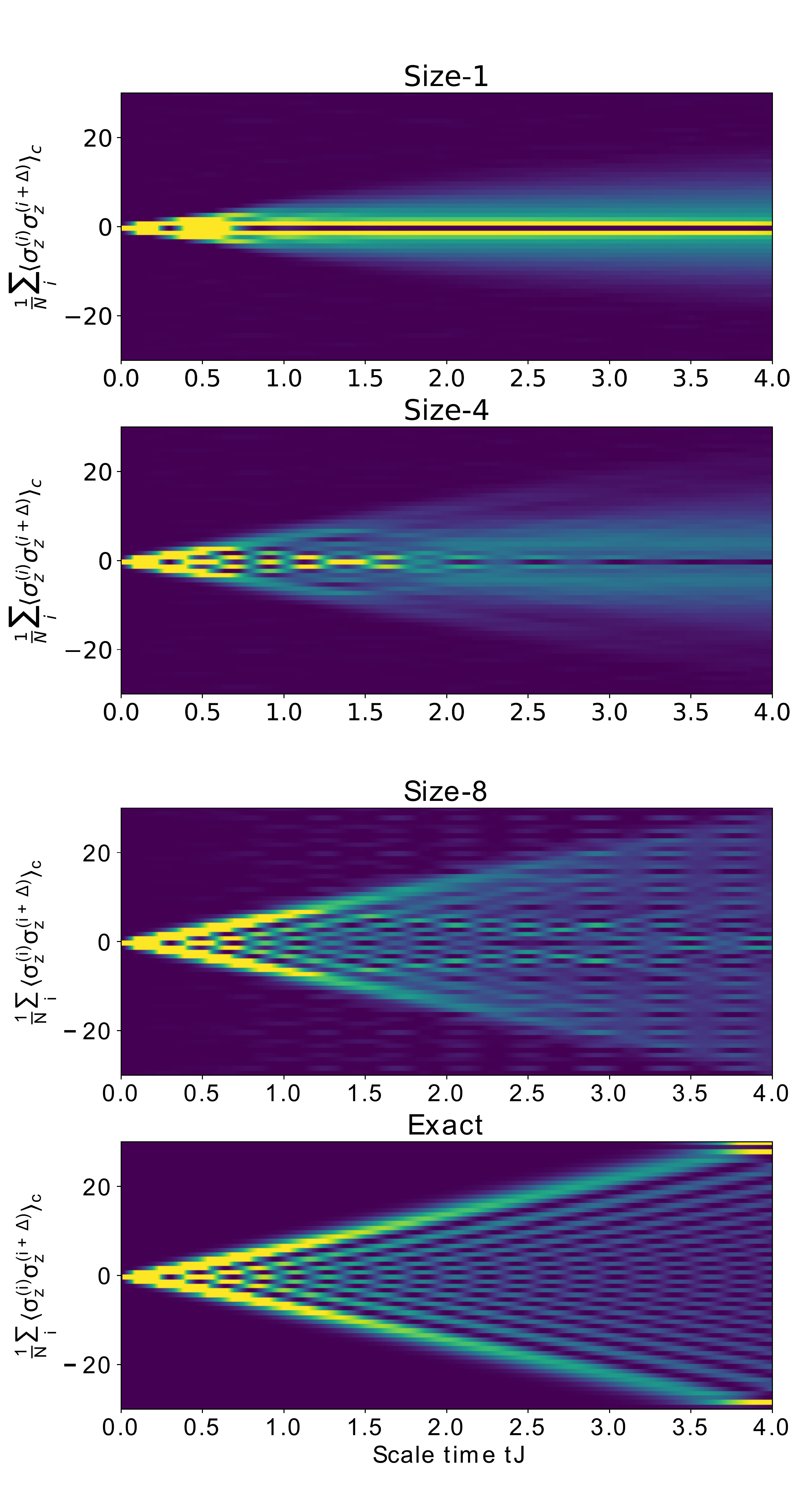}
		\caption{\textbf{
		The connected correlator for the 1d XY spin chain} as outlined in Eqs.~\eqref{eq:ZZ_connected_correlator}, \eqref{eq:XY_hamiltonian}. Ballistic growth of entanglement is well encoded in the correlations of the semiclassical phase space, improving with cluster size. Color encodes strength of the expectation value of the connected correlator $\langle \hat \sigma_z^{(i)} \sigma_z^{(i+\Delta)}\rangle_{c}$.}\label{fig:connected_correlator}
	\end{figure}

Using Eq.~\eqref{eq:operator_observables} CTWA also allows one to compute various local and non-local correlation functions. Suppose we want compute the equal time connected correlation function of $z$-magnetization between sites $i$ and $j$. If these sites belong to different clusters  then
\begin{eqnarray}\label{eq:ZZ_connected_correlator}
&&\langle \hat{\sigma}^{(i)}_z(t)\hat{\sigma}^{(j)}_z(t)\rangle -\langle\hat{\sigma}^{(i)}_z(t)\rangle\langle\hat{\sigma}^{(i+\Delta)}_z(t)\rangle\\
&&=\langle \hat X_\alpha^{(i')}(t) \hat X_\beta^{(j')}(t) \rangle-\langle \hat X_\alpha^{(i')}(t) \rangle \langle \hat X_\beta^{(j')}(t) \rangle\\ 
=&&\overline{x_\alpha^{i'}(t)x_\beta^{j'}(t)} - \overline{x_\alpha^{i'}(t)}\;\overline{x_\beta^{j'}(t)}.
\end{eqnarray}
where $i'$ ($j'$) is the cluster containing site $i$ ($j$) and $\alpha$ and $\beta$ are the indices corresponding to representation of Pauli matrices through the cluster basic operators (in general the observables of interest can be represented through linear combinations of the basis operators). If the sites $i$ and $j$ belong to the same cluster $i'$ then computing the correlation function is even simpler as the whole product $\langle \hat{\sigma}^{(i)}_z(t)\hat{\sigma}^{(j)}_z(t)\rangle$ is represented through a linear combination of the basis operators $\hat X_\alpha^{(i')}$.

These correlation functions are sensitive to entanglement/mutual information between different sites and like the entanglement entropy can not be captured within mean-field approximations as they arise, within CTWA, solely due to the initial fluctuations which propagate differently in space-time. In Fig.~\ref{fig:connected_correlator} we show this correlation function for the periodic 64-site XY model at zero disorder:

\begin{equation}\label{eq:XY_hamiltonian}
\hat H = \sum_{i=0}^{64}\hat \sigma_x^{(i)}\hat{\sigma}_x^{(i+1)} +\hat \sigma_y^{(i)}\hat{\sigma}_y^{(i+1)}.
\end{equation}

This Hamiltonian is well-known to be integrable and mappable to free fermions. In fact under an alternate operator basis introduced in Ref.~\onlinecite{Davidson2017} the TWA dynamics for this model become exact. This emphasizes that basis choice $\{\hat X^{{i'}}_\alpha\}$ can be important for the accuracy of CTWA. The connected correlator clearly approaches the exact result for larger cluster sizes, and is nonzero even far outside the span of a cluster, indicating that entanglement information can be well encoded in the phase space.

\section{CTWA Details}
\label{sec:GaussCTWA}
In this section we elaborate some details of CTWA, starting from how one can define and sample the Gaussian Wigner function as well as detail implementing efficient CTWA dynamics. We will also provide some additional justifications for the choice of initial conditions and comment on connections between operator and Schwinger boson CTWA schemes.

\subsection{The Gaussian Wigner function for operator CTWA.}\label{sec:gauss_wigner_function_details}

We formally introduced the Gaussian Wigner function in Eqs.~\eqref{eq:gaussian_wigner_function} and~\eqref{eq:gaussian_sampling}. For simplicity we will consider here, as before, only product state initial density matrices such that the Wigner function also factorizes. For this reason we can drop the cluster index in the notation Wigner function in this section. For completeness we repeat its definition here:
\begin{eqnarray}
\label{eq:gaussian_wigner_function_ap}
	&&W(\{x\})={1\over Z}\exp\left[(x_\alpha-\rho_\alpha) \Sigma_{\alpha\beta} (x_\beta-\rho_\beta)\right],\\
	&&{\rm Tr}\big[\hat \rho \hat X_\beta\big]=\int \prod_\alpha dx_\alpha\, x_\beta\, W(\{x\}),\nonumber\\
	&&{\rm Tr}\big[\hat \rho (\hat X_\beta\hat X_\gamma+\hat X_\gamma\hat X_\beta)\big]=2\int \prod_\alpha dx_\alpha\, x_\beta x_\gamma W(\{x\}).\nonumber
\end{eqnarray}

\begin{figure}
	\centering
	\includegraphics[width=\linewidth]{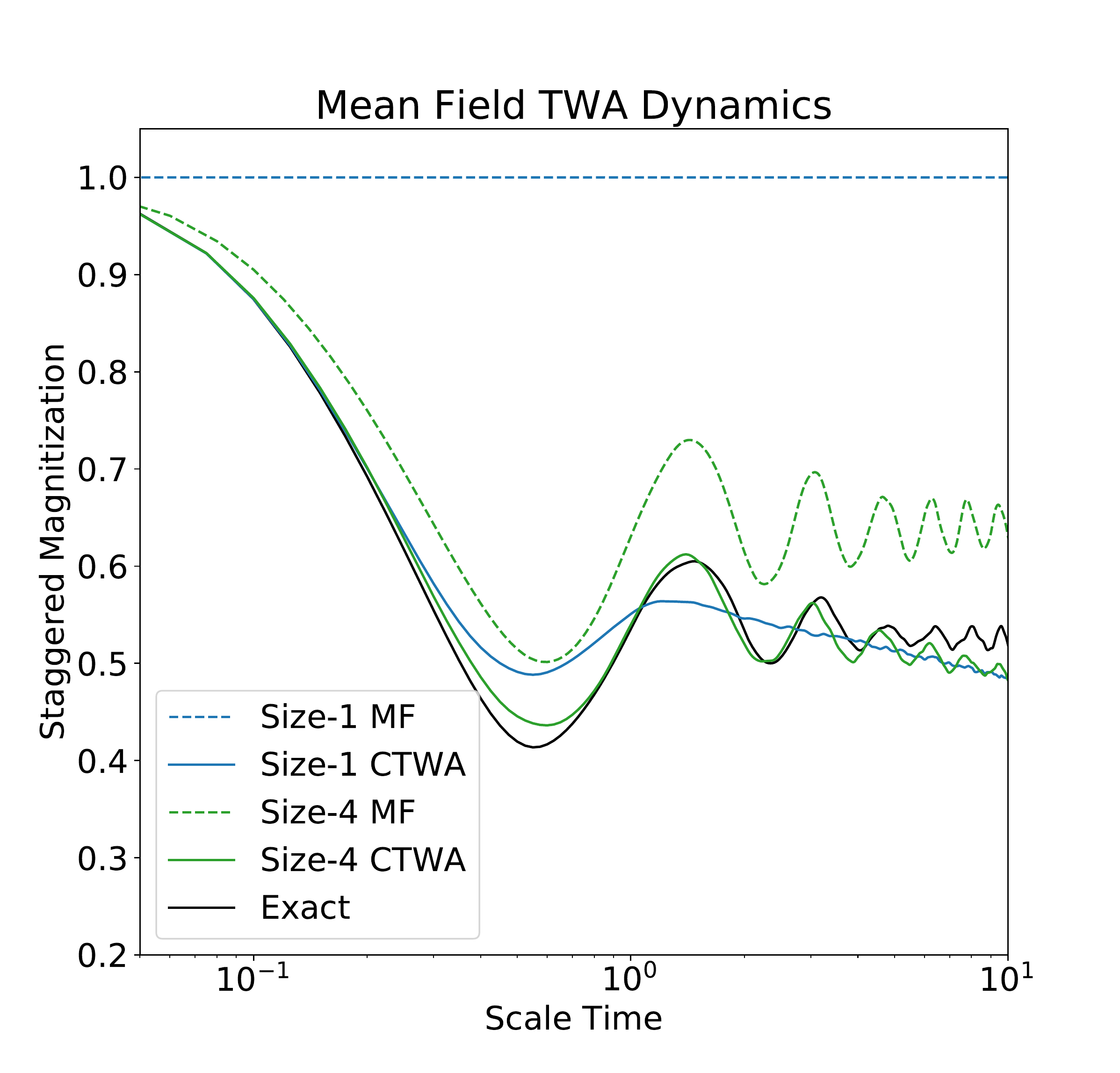}
	\caption{\textbf{Mean Field dynamics for the Disordered Heisenberg model}
		 \eqref{eq:heisenberg_hamiltonian}, neel initial condition, and staggered magnitization observable \eqref{eq:neelstate}, averaged over disorder $h_z=10$. Clearly for increased cluster size dynamics approach exact results, and fluctuations are required to reproduce dynamics at short times.}\label{fig:flucsizes}
\end{figure}

While our Gaussian choice for the Wigner function looks a little ad-hoc, there are several reasons justifying such choice (beyond the fact that it is easy to sample):

\begin{itemize}

\item Standard TWA, which is formally controlled by $\hbar$  or $1/S$ in the spin models with large $S$ corresponding to the classical limit is only accurate to the order $\hbar^2$ or $1/S^2$ (see Ref.~\onlinecite{Polkovnikov2010}). For this reason one can approximate the Wigner function to the same order without the loss of accuracy. For pure spin states, e.g. states polarized along the $z$-axis, higher order cumulants are suppressed~\footnote{For standard $SU(2)$ spins this fact trivially follows that in the classical limit the polarized spin is represented by the non-fluctuating $\delta$-function distribution and fluctuations are quantum coming from non-commutativity of spin components. Same arguments can be applied to higher-dimensional spins.} by powers of $1/S$ and so it is only necessary to correctly describe first two cumulants: the mean and the variance. Any mixed state can be represented as a statistical mixture of pure states, so the associated Wigner function can be approximated as a sum of Gaussian distributions with non-negative weights. Interestingly in Ref.~\onlinecite{Davidson2015} it was observed numerically that in the $SU(3)$ case the Gaussian Wigner function results in a slightly better approximation to dynamics than the exact Wigner function. On passing we note the Gaussian approximations to the density matrices representing quantum statistical ensembles close to the classical limit also correctly capture leading quantum corrections to the corresponding classical probability distributions~\cite{Kim2017}.

\item Fixing second moments of the Wigner Function ensures that CTWA is asymptotically exact at short times up to the order $\mathcal O(t^2)$, which is the same accuracy as normal TWA. Similarly, the mean-field approximation, which sets all fluctuations to zero, is only accurate to the linear order in time. This can be seen by Taylor expanding the evolution of some operator $\hat O$. In linear order the response will be determined by the initial expectation value of the commutator $\langle \psi_0| [\hat O,\hat H]|\psi_0\rangle$, which is by construction is linear in cluster variables for any product state $|\psi_0\rangle$. In the next $t^2$ order, the response will involve expectations values of the higher order commutators like $\langle \psi_0| [[\hat O,\hat H],\hat H]|\psi_0\rangle$. It is straightforward to check that such double commutators will involve only linear or quadratic terms in the cluster operators, which are again guaranteed to be exactly reproduced by the Gaussian Wigner function. We point in this respect that absence of fluctuations in initial conditions in dynamic mean-field approximations generally leads to the mistake in the expectation value of the double commutator and hence leads only to accuracy up to the linear order in $t$. Therefore initial quantum fluctuations encoded in the width of the Wigner function guarantee the correct short time dynamics. Beyond the order $t^2$ there are generally quantum jump contributions to the dynamics, which go beyond TWA~\cite{Polkovnikov2010}, and so going beyond the Gaussian approximation of the Wigner function will generally not improve the accuracy of CTWA, at least at short times.

The importance of fluctuations is shown in figure \ref{fig:flucsizes}, where dynamics of the disordered Heisenberg model are compared to that of mean-field CTWA. Clearly, without fluctuations, the second derivative of the observable (eg, $\mathcal O(t^2)$) is not correctly approximated.

\item The exact Wigner function, which can be e.g. obtained from a Schwinger boson representation of the basis operators $\hat X_\alpha$ has a huge (exponential in the cluster size) amount of noise in the system. It follows from the fact the each unoccupied Schwinger boson has a $1/2$ quantum noise, and the number of Schwinger bosons is equal to the Hilbert space dimension of the cluster. This huge noise is both unphysical and would make long time dynamics uncontrollable leading to the same issues which traditional TWA faces due to diverging ultraviolet noise coming from zero point fluctuations of vacuum modes~\cite{Blakie2008}. The Gaussian Wigner function we are using here does not have this problem and the quantum noise does not diverge with the increasing cluster size. Empirically, we find that this not only allows us to do controlled accurate simulations of quantum dynamics, but in many cases CTWA with the Gaussian Wigner function accurately predicts both the short time dynamics and the long time thermalization.

\end{itemize}


\subsection{Alternate Operator Basis for Operator CTWA}\label{sec:alternate_basis_CTWA}
In the previous sections we use an operator basis of products of Pauli matrices (see Sec.~\ref{sec:operatorclusterTWA}) because it has transparent physical interpretation. However for sampling purposes it is more convenient to use the basis of rank $1$ matrices $\hat Y_{nm}$, $n= 0,1\dots D-1,\; m=1,2\dots n$, defined as:
\begin{align*}
m=n \;&\quad \hat Y_{nn}=|n\rangle\langle n|,\\
m<n\; &\quad \hat Y_{nm}^{(+)}={1\over \sqrt{2}}(|n\rangle\langle m|+|m\rangle\langle n|),\\
&\quad Y_{nm}^{(-)}={i\over \sqrt{2}}(|n\rangle\langle m|-|m\rangle\langle n|),
\end{align*}
where $|n\rangle$ is the $n$-th state in the standard product basis of spins polarized in the $z$-direction. As in the main text we use the convention that $|0\rangle=|\uparrow\uparrow\dots\uparrow\rangle$, $|1\rangle=|\downarrow\uparrow\dots\uparrow\rangle,\;\dots$,
These operators are explicitly Hermitian clearly forming a complete operator basis. The structure constants for these operators are easy to find from the commutation relations, e.g.
\begin{multline}
[\hat Y_{nm}^{(+)}, \hat Y_{pq}^{(-)}]=\\
{i\over \sqrt{2}} \left(
\delta_{mp} \hat Y_{nq}^{(+)}+\delta_{np} \hat Y_{mq}^{(+)}-\delta_{nq} \hat Y_{mp}^{(+)}-\delta_{mq} \hat Y_{np}^{(+)} \right).
\end{multline}

As is shown in section \ref{sec:general_ICs}, any product state wave function initial condition can be mapped from an initially Z-polarized state $|\psi\rangle = |0\rangle$. Thus, defining the Wigner function for this state is sufficient to define any Wigner function. For this state, the correct sampling of the phase space coordinates $y_{nm}$ is given by:

\be
y_{00}=1,\; y_{n0}^{(+)}={\delta_n\over \sqrt{2}},\; y_{n0}^{(-)}={\sigma_n\over \sqrt{2}},\; y_{nm}=0,\quad m\neq 0.
\label{eq:sampling_y}
\ee
Here $\delta_n$ and $\sigma_n$ are independent Gaussian random variables with zero mean and unit variance. The proof of Eq.~\eqref{eq:sampling_y} is rather straightforward. First of all let us observe that in the state $|\psi\rangle=|0\rangle$ only the operator $\hat Y_{00}$ has a non-vanishing expectation value
\[
\langle 0|\hat Y_{00}|0\rangle=\langle 0|0\rangle\langle 0|0\rangle=1=\overline{y_{00}}.
\]
This expectation value is obviously reproduced by the chosen distribution function
\begin{align}
\langle 0|\hat Y_{00}^2|0\rangle=1 \quad & \quad \overline{y_{00}^2}=1 \nonumber \\
\langle 0| \{\hat Y_{0p}^{(+)} ,Y_{0q}^{(+)}\}_+|0\rangle=\delta_{pq} \quad & \quad 2\overline{y_{0p}^{(+)}y_{0q}^{(+)}}=\delta_{pq}
\label{eq:dist_Y} \\
\langle 0| \{\hat Y_{0p}^{(-)} ,Y_{0q}^{(-)}\}_+|0\rangle= \delta_{pq} \quad & \quad 2\overline{y_{0p}^{(-)}y_{0q}^{(-)}}= \delta_{pq}.\nonumber
\end{align}
It is easy to verify that the expectation value of all other anticommutators vanish; likewise all other correlations of the associated phase space variables vanish too. 

\subsection{General initial conditions for Operator CTWA}\label{sec:general_ICs}

As mentioned above, any product state within a cluster can be mapped from the Z-polarized state via some rotation. This may generally be done via a unitary coordinate rotation $x_\alpha \to U_{\alpha\beta}x_\beta$. Sometimes it is easier to to do instead a local (intra-cluster) Hamiltonian evolution, where CTWA is exact. For example, for the state $|\downarrow\uparrow\uparrow\rangle$ we can sample a point from the Wigner function corresponding to the state $|0\rangle=|\uparrow\uparrow\uparrow\rangle$, then evolve the Hamiltonian $\hat H=\sigma_x^{(0)}$ for time $t=\pi$. It is easy to prove that this way of sampling the initial conditions is equivalent to sampling from the Gaussian Wigner function corresponding to the state $|\downarrow\uparrow\uparrow\rangle$.  In this way, the initial states mentioned in the paper can be readily generated. E.g. for the Neel state one can first sample the state $|0\rangle$ and then apply the Hamiltonian $\hat H = \sum_{i\text{ even}}\hat \sigma_x^{(i)}$ for time $t=\pi$. The Wigner function for a random state of the Ising example was produced in a similar fashion by applying the Hamiltonian $\hat H = \sum_i h_x^i\hat \sigma_x^{(i)}+h_y^i\hat \sigma_y^{(i)}+h_z^i\hat \sigma_z^{(i)}$ for $t=\pi$ with the randomly chosen magnetic fields $\vec h$ and so on.

\subsection{Dimensional Reduction for Operator CTWA}\label{sec:dim_red_CTWA}

Intriguingly, for a Gaussian Wigner and pure initial state, the complexity of the system can be reduced from $D^2$ to $2D$. To see this let us define a Hermitian matrix $y$ consisting of elements $y_{nn}$ on diagonal , $y_{mn}=1/\sqrt{2} (y_{mn}^{(+)}-i y_{mn}^{(-)})$ on the upper triangular and $y_{nm}=1/\sqrt{2} (y_{mn}^{(+)}+i y_{mn}^{(-)})$ on the lower triangular. From Eq.~\eqref{eq:dist_Y} it follows that the only nonzero matrix elements of $y$ are $y_{00}=1$, $y_{0n}$ and $y_{n0}$:
\begin{equation}
y=\left(
\begin{array}{ccccc}
1 & {\delta_1-i\sigma_1\over 2} &{\delta_2-i\sigma_2\over 2} & \dots & {\delta_D-i\sigma_D\over 2}\\
{\delta_1+ i\sigma_1\over 2} & 0 & 0 & \dots &0\\
{\delta_2+ i\sigma_2\over 2} & 0 & 0 & \dots &0\\
\vdots & \vdots & \vdots &\ddots &\vdots\\
{\delta_D+ i\sigma_D\over 2} & 0 & 0 & \dots &0
\end{array}
\right).
\label{eq:ymatrix}
\end{equation}

This matrix has only 2 non-zero eigenvalues $\lambda_{\pm}$, allowing us to express it as
\begin{equation}
y=\lambda_+ |\psi_+\rangle \langle \psi_+| +\lambda_- |\psi_-\rangle\langle \psi_-|,
\label{eq:ypm}
\end{equation}
where $|\psi_\pm\rangle$ are the eigenvectors of $y$ corresponding to the two non-zero eigenvalues $\lambda_\pm$, which  can be obtained from the characteristic polynomial:
\begin{equation}
f(\lambda)=(-\lambda)^{D-2} \left( (1-\lambda)(-\lambda)-\frac{1}{4}\sum_{i=1}^{D-1} (\delta_i^2+\sigma_i^2) \right).
\end{equation}
Then the only two non-zero eigenvalues are
\begin{equation}
\lambda_{\pm}=\frac{1}{2} \pm \frac{1}{2} \sqrt{1+\sum_{i=1}^{D-1} (\delta_i^2+\sigma_i^2) }.
\end{equation}
The eigenvalues can be intepreted as quasi-probabilities, since $\lambda_++\lambda_-=1$. Note that, one the eigenvalues is always positive and the other is always negative. Alternatively one can think of them as the components of an auxiliary spin 1/2 degree of freedom per cluster. Note that the eigenvalue only depends on the sum of all the $\delta_i$ and $\sigma_i$. Given that they are Gaussian i.i.d. variables, the eigenvalues converge to 
\begin{equation}
\lambda_\pm\approx \frac{1}{2} \pm \frac{1}{2} \sqrt{2D-1}\approx \frac{1}{2} \pm \sqrt{\frac{D}{2}},
\end{equation}
in the large $D$ limit. Fluctuations in the eigenvalues from realization to realization are subleading and of order $O(D^{1/4})$. The eigenvectors on the other hand become,
\begin{eqnarray}
&&\left<0|\psi_\pm\right>=\sqrt{\frac{\lambda_\pm^2}{\lambda_\pm^2+1/2}}, \nonumber \\
&&\left<i|\psi_\pm\right>=\sqrt{\frac{\lambda^2_\pm }{\lambda_\pm^2+1/2}} \frac{\delta_i+i\sigma_i}{2\lambda_\pm} \quad\forall\, i> 0.
\end{eqnarray}
For sufficiently large clusters this behaves as
\begin{eqnarray}
&&\left<0|\psi_\pm\right>\approx 1, \nonumber \\
&&\left<i|\psi_\pm\right>\approx \pm  \frac{\delta_i+i\sigma_i}{\sqrt{2D}} \quad\forall\, i> 0.
\end{eqnarray}
Both eigenvectors of $y$ are therefore the initial state $\left| 0\right>$ supplemented by exponentially small noise in cluster size. This is the primary reason why CTWA doesn't suffer from any problematic noise accumulation. Indeed, from~\eqref{eq:ymatrix} it might have appeared that CTWA is adding an exponential amount of noise to the initial conditions. However, the above analysis shows that it actually results in exponentially large eigenvalues with exponential suppression of the fluctuations in each of the eigenvectors. It can be shown using the equations of motion (\ref{eq:twa_eq0}) that these eigenvalues are conserved in time, and thus the noise remains confined at {\em all} times.

We can use conservation of $\lambda_\pm$ to further reduce the number of equations of motion for the phase space variables from naive $\mathcal O(ND^2)$, where $N$ is the number of clusters to $\mathcal O(ND)$. For example, if we have two clusters denoted by $L$ and $R$  and a Hamiltonian:
\[
\hat H=\hat H_{L}+\hat H_{R}+J\hat V_L \hat V_R
\]
then equations of motion for the variables $y_{nm}$ are equivalent to equations of motion for the states $|\psi_\pm \rangle$ with each cluster. Thus:
\begin{multline}
i\partial_t |\psi^L_\pm(t)\rangle=\hat H_L |\psi_\pm^L(t) \rangle+ J \hat V_L |\psi_\pm^L(t)\rangle\times\\
\times \left( \lambda_+^{(R)}  \langle \psi_+^R(t) | \hat V_R
|\psi^R_+(t)\rangle+ \lambda_-^{(R)}  \langle \psi_-^R(t) | \hat V_R |\psi_-^R(t)\rangle\right),
\label{eq:evolution_vectors}
\end{multline}
and similarly for the right two eigenvectors. At the end of the evolution the phase space variables $y_{nm}(t)$ for each for the clusters are obtained from the state vectors $|\psi_\pm^{(L,R)}\rangle$ according to Eq.~\eqref{eq:ypm}. Let us comment that clearly this way of simulating dynamics of cluster variables bears many parallels with Schwinger boson CTWA simulations, which we will discuss in some detail below.

\subsection{Gaussian Wigner function for Schwinger bosons}\label{sec:SB_TWA_details}

In the previous section we discussed that the $4^N$ complexity of the cluster variables can be reduced to $2^N$ complexity by the appropriate sampling of the matrices by vectors and evolving these vectors like wave functions (c.f. Eq.~\eqref{eq:ypm} and the following discussion). It is interesting that the two eigenvalues $\lambda_\pm$ appearing in this sampling (c.f. Eqs.~\eqref{eq:ypm} and~\eqref{eq:evolution_vectors}) can be interpreted as discrete quasi-probabilities as they do not evolve in time. As we discussed $\lambda_-<0$ so we are effectively dealing with negative quasi-probabilities (alternatively the two numbers $\lambda_+$ and $\lambda_-$ represent couplings to $z$-components of one auxiliary spin per cluster, which does not evolve in time). This of course does not create a severe sign problem because we only have a single negative number per cluster containing exponentially many variables.

Instead of this sampling of the operator variables, one can directly work with Schwinger bosons (see Sec.~\ref{sec:SB_TWA}). There we are dealing with $\mathcal O(2^N)$ independent complex numbers from the beginning. It is interesting that reproducing all fluctuations of the basis operators in the language of Schwinger bosons also requires a single negative quasi-probability per cluster. This is mathematically related to the fact that we are dealing with the Fock state containing one boson. 

In the Schwinger boson representation the basis operators are expressed through the bilinears  of creation and annihilation operators $\hat b_a^\dagger,\, \hat b_a$ and the corresponding phase space variables are represented similarly through the complex bosonic fields $b_a^\ast, b_a$ satisfying canonical poisson bracket relations $\{b_a^\ast, b_b\}=-i\delta_{ab}$ (c.f. Refs.~\onlinecite{Polkovnikov2010, Davidson2015}):
\begin{equation}
	\hat X_\alpha=\hat b_a^\dagger T_\alpha^{ab} \hat b_b,\quad x_\alpha=b_a^\ast T_{\alpha}^{ab} b_b.
	\label{eq:def_schwinger_bosons}
	\end{equation}
Here $T_\alpha$ are the $D$-dimensional matrices realizing the $SU(D)$ algebra with the structure constants $f_{\alpha\beta\gamma}$:
\[
[T_\alpha, T_\beta]=i f_{\alpha\beta\gamma} T_\gamma.
\] 
In the fundamental (spin one/half) representation of the group we work with the total number of Schwinger bosons should be one: 
\[
\hat n=\sum_a \hat b_a^\dagger \hat b_a=\hat {\bf I}.
\] 
Because $\hat n$ is conserved in time for any Hamiltonian expressed through the spin operators $\hat X_\alpha$ one only needs to impose this constraint at initial conditions. Physically the Schwinger boson simply labels the expansion of the wave function of the system in some basis, e.g. given by Eq.~\eqref{eq:cluster_basis} and the matrices $T_\alpha$ are simply given by the matrix elements of the corresponding basis operators $\hat X_\alpha$ in this basis: $T_\alpha^{ab}=\langle a| \hat X_\alpha |b\rangle$. 

As in the case with clusters in order to reproduce correctly initial conditions for any pure state it is sufficient to reproduce them for the state $|0\rangle=|\uparrow\uparrow\dots \uparrow\rangle$, which is in the language of Schwinger bosons is represented as $|100\dots 0\rangle$. In principle one can use the exact Wigner function for this state, which is given by the product of the Wigner function of the Fock state with one boson for the boson $b_0$ and of the Wigner functions corresponding to the vacuum state for all other bosons:
\begin{equation}
W_{sb}(\vec b,\vec b^*)= L_1(4|b_0|^2)\prod_{a=0}^{D-1} \mathrm e^{-2|b_a|^2}
\end{equation}
This choice of the Wigner function is very inconvenient for large clusters as it leads to exponentially hard sampling and very large errors in the long time limit. The reason behind is that each vacuum mode contributes noise of the order of unity to any basis operator and hence the total number of noise in the system scales exponentially with the cluster size $N$. Moreover, we checked numerically that even if we can circumvent the sampling problem for moderate size clusters, this Wigner function leads to large, essentially uncontrolled, mistakes at long times. The reason is that under nonlinear TWA equations this noise describing virtual vacuum excitation can deplete leading to negative occupation numbers of empty modes. In fact, this problem is very well known in standard bosonic TWA through spurious dependence of long time TWA dynamics on ultriviolet cutoff~\cite{Blakie2008}.

We find that it is much more convenient and accurate to use a phenomenological Wigner function, which fixes first and second moments of the basis operators in the initial state:
\begin{eqnarray}
	\langle \hat{X}_\alpha \rangle &=& \sum_{ab} T_\alpha^{ab}\int d\vec bd\vec b^\ast W(\vec{b},\vec b^*)b_a^\ast   b_b,\label{eq:SB_wignerfunc}\\
\left<\{\hat X_\alpha,\hat{X}_\beta\}_+\right> &=& 2\sum_{ab,cd} T^{ab}_\alpha T^{cd}_\beta\int d\vec b d\vec b^\ast W(\vec{b},\vec b^\ast)b_a^\ast b_b b_c^\ast b_d.\nonumber
	\end{eqnarray}
It is easy to check that for the state $|100\dots 0\rangle$ the Wigner function like in the exact case factorizes into the product of $w_0(b_0,b_0^\ast)$ corresponding to the occupied boson and the product of identical Wigner functions describing empty modes:
\begin{equation}
W(\vec b,\vec b^\ast)=w_0(b_0,b_0^\ast)\prod_{a=1}^{D-1} w_1(b_a^\ast, b_a).
\end{equation}
In order to fix the correct expectation values and fluctuations of the basis operators in the $|\uparrow\dots \uparrow\rangle$ state it is sufficient to consider only one diagonal and one off-diagonal operator, e.g. $\hat \sigma_z^{(1)}$ and $\hat \sigma_x^{(1)}$. It is easy to see that all other diagonal and off-diagonal operators are obtained from these two by permutations of Schwinger bosons $b_1,\dots b_{D-1}$. Because those have identical distributions such permutations do not affect any expectation values. Then the four non-equivalent expectation values we need to reproduce are
\begin{eqnarray*}
1&=&\langle \sigma_z^{(1)}\rangle=\overline{|b_0|^2-|b_1|^2+|b_2|^2-|b_3|^2+\dots}=\overline{|b_0|^2} -\overline{|b_1|^2},\\
0 &=&\langle \sigma_x^{(1)}\rangle=\overline{ b_0^\ast b_1+b_1^\ast b_0+b_2^\ast b_3+b_3^\ast b_2+\dots},\\
1 &=& \langle \left(\sigma_z^{(1)}\right)^2\rangle=\overline{|b_0|^4}+(D-1) \overline{|b_1|^4}-2\overline{|b_0|^2}\,\overline{|b_1|^2}-D (\overline{|b_1|^2})^2,\\
1 &=& \langle \left(\sigma_x^{(1)}\right)^2\rangle=2\overline{|b_0|^2}\,\overline{|b_1|^2}+(D-2) (\overline{|b_1|^2})^2.
\end{eqnarray*}
The second equation is trivially satisfied because for the Fock state the Wigner function cannot depend on the phases of the bosons. Combining the first and the last equation and assuming that for the empty modes ($b_1$) we can use the Gaussian ansatz such that $\overline{|b_1|^4}=2(\overline{|b_1|^2})^2\equiv 2 \sigma_1^4$ we find
\be
D\sigma_1^4+2\sigma_1^2-1=0\;\Rightarrow\; \sigma_1^2={\sqrt{1+D}-1\over D}.
\label{eq:noise_SB_twa}
\ee
exactly as in Eq.~\eqref{eq:SB_wignerfunction}. Plugging this into the remaining two equations we find
\begin{eqnarray}
\label{eq:SB_moments}
\overline{|b_0|^2}=1+\sigma_1^2, \nonumber \\
\overline{|b_0|^4}=2-2(D-2)\sigma_1^4.
\end{eqnarray}

It is easy to see that this condition can not be satisfied with a positive probability distribution for any $D\geq 2$. We find empirically in all examples we simulated that one can simply suppress fluctuations in $\overline{|b_0|^2}$ all the way to zero like we did in Eq.~\eqref{eq:SB_wignerfunction} without affecting accuracy of TWA both at short and at long times. This means that Eq.~\eqref{eq:SB_wignerfunc} is not fully obeyed. In particular, the expectation value of $\hat \sigma_z^2$ is not reproduced correctly by the simplified Wigner function. However, this expectation value does not enter the linear response and thus does not affect the accuracy of the short time dynamics. It is equally unimportant in the long time limit. One can check that in order to satisfy Eq.~\eqref{eq:SB_moments} one can choose an appropriate discrete probability distribution for $|b_0|^2$ characterized by one positive $\lambda_+$ and one negative $\lambda_-$ probabilities in direct analogy with eigenvalues $\lambda_\pm$ discussed in the previous section. Interestingly the expectation value of the identity operator $\langle 0|\hat I|0\rangle=\sum_j \overline{|b_j|^2}$ scales as $\sqrt{D}$ in the large $D$ limit, which implies that one has to be cautious in literary identifying Schwinger bosons with the components of the wave function. However, the identity operator (equal to the total Schwinger boson number operator) is conserved in time and does not affect expectation values of any traceless operators we are interested in.

As a final comment, remarkably the Schwinger boson Wigner function with the noise set by Eq.~\eqref{eq:noise_SB_twa} is much more physical than the exact Wigner function. At larger $D$ we occupy each empty mode with $1/\sqrt{D}$ noise such that the total amount of noise in the system is always of the order of one, independent of the cluster size, as in the operator CTWA method.

	\section{Conclusions}
	
In this work, we developed a general method for simulating dynamics of interacting quantum systems using cluster semiclassical (truncated Wigner) approximation, which we termed CTWA. The method is based on first including various local correlations within a cluster as additional degrees of freedom effectively increasing the phase space dimensionality and then utilizing the TWA in this higher dimensional space. As particular illustrations of CTWA we focus on quantum spin one-half chains, which we split into clusters of size $N$. Each cluster is spanned by $D^2=4^N$ basis operators forming a closed $SU(D)$ algebra.  These operators are then treated as high-dimensional spin components. The number of degrees of freedom can be further reduced to $D=2^N$ either using the Schwinger boson representation of the operators or using other decompositions of operators into products of $D$-dimensional vectors, which can be alternatively sampled and propagated. Formally CTWA equations are identical to the mean field equations for the density matrix if we use operators as classical variables and to the mean field equations for the wave function if we use the Schwinger Boson representation.

A crucial difference between CTWA and mean field is the presence of fluctuations in initial conditions encoded in the initial Wigner function. These fluctuations allow CTWA to correctly reproduce short time dynamics to a higher order in time than the mean field. They also allow capturing entanglement between subsystems and accurately reproducing local and non-local connected correlation functions of various observables via the classical mutual information in phase space, which are beyond mean field approximations. To avoid the unphsyical exponential in the cluster size noise in initial conditions we use the Gaussian approximation to the Wigner function. The parameters of the Gaussian are chosen such that the Gaussian Wigner function reproduces the expectation values and fluctuations of the basis operators in the initial state.

We demonstrate the method simulating dynamics in various strongly coupled spin-chains with and without disorder. As computational difficulty scales linearly in system size, we simulate both small 1D systems (to allow for comparison with exact diagonalization) as well as in larger systems. In all the cases CTWA predictions approach exact results with an increasing cluster size. In thermalizing regimes we find numerically that CTWA accurately captures the long time thermal phase and the hydrodynamic (diffusive) approach to the equilibrium. In the MBL regime CTWA reproduces the intermediate time localization plateau but eventually predicts thermalization, with thermalization time increasing with the cluster size. We also demonstrate that CTWA predictions for entanglement closely resemble exact results.
	
While in this work we focused on a particular cluster choice of phase space variables we note that the method can be applied to any operator basis  $\{\hat X_\alpha\}$ which forms a closed algebra, as long as the Hamiltonian and observables of interest can be expressed through sums and products of these basis operators. The basis operators then map to phase space variables, which evolve in time according to the corresponding classical Hamiltonian equations of motion. Note that as the operator algebra is invariant under any unitary transformation, one can take some fixed operator basis unitarily evolve \{$\hat X_\alpha\}\to \{\hat U^\dagger \hat X_\alpha \hat U\}$ to form a new (nonlocal) operator basis and then apply CTWA. We expect that in this way one can further improve accuracy of the method. One can also generalize CTWA to the situations where we deal with the approximate operator basis, which does not form a closed algebra, e.g. a Matrix Product State basis like the one used in Refs.~\onlinecite{Kramer2008,Leviatan2017, Shi2017} in the context of time-dependent variational method. Technically this can be done by using the second of Eqs.~\eqref{eq:structure_constants} defining the structure constants. In this way one effectively projects the commutator between any two basis operators to the subspace spanned by these operators. Finally one can combine the cluster CTWA with the fermionic TWA developed in Ref.~\onlinecite{Davidson2017} to describe dynamics of strongly interacting fermionic systems.

%
%
	
	\acknowledgements  We would like to thank Shainen Davidson for collaboration on early stages of this work and on many valuable discussions. J.W. and A.P. were partially supported by NSF DMR-1506340 and AFOSR FA9550-16- 1-0334. D.S. acknowledges support from the FWO as post-doctoral fellow of the Research Foundation -- Flanders and CMTV.

\appendix

\bibliographystyle{apsrev4-1}

\bibliography{CTWA_refs}

\end{document}